\documentclass[10pt,journal,compsoc]{IEEEtran}

\ifCLASSOPTIONcompsoc
  \usepackage[nocompress]{cite}
\else
  \usepackage{cite}
\fi

\ifCLASSINFOpdf
  \usepackage[pdftex]{graphicx}
  \DeclareGraphicsExtensions{.pdf,.jpg,.jpeg,.png}
\else
  \usepackage[dvips]{graphicx}
  DeclareGraphicsExtensions{.eps}
\fi

\usepackage{amsmath}

\usepackage{amssymb}%
\usepackage{pifont}%

\usepackage{multirow}

\usepackage[T1]{fontenc}
\usepackage{subcaption}
\usepackage[numbers]{natbib}
\usepackage{bm}
\usepackage{changes}
\usepackage{todonotes}
\usepackage{xspace}
\usepackage{booktabs}
\usepackage[colorlinks=true, allcolors=black]{hyperref}

\newcommand\cmr[1]{\textcolor{black}{#1}}

\newcommand{\datasetName}{Visualisation Recallability Dataset\xspace}
\newcommand{\datasetNameShort}{VisRecall\xspace}

\newcommand{\methodNameLong}{Recallability Network\xspace}
\newcommand{\methodNameShort}{RecallNet\xspace}

\begin{document}

\title{\datasetNameShort: Quantifying Information Visualisation Recallability via Question Answering}

\author{Yao~Wang,~Chuhan~Jiao,~Mihai~B\^{a}ce,~and~Andreas~Bulling%
\IEEEcompsocitemizethanks{\IEEEcompsocthanksitem 
Yao~Wang,~Mihai~B\^{a}ce,~and~Andreas~Bulling are with the Institute for Visualisation and Interactive Systems, University of Stuttgart, Germany, \protect E-mail: \{yao.wang, mihai.bace, andreas.bulling\}@vis.uni-stuttgart.de.%
\IEEEcompsocthanksitem Chuhan~Jiao is with Aalto University, E-mail: chuhan.jiao@aalto.fi
\IEEEcompsocthanksitem Yao Wang is the corresponding author.}%
%\thanks{Manuscript received xx xx, 2021; revised xx xx, 202x.}
}

\markboth{}%
%\markboth{IEEE Transactions on Vizualisation and Computer Graphics}%
{Wang \MakeLowercase{\textit{et al.}}: VisRecall: Quantifying information Visualisation Recallability via Question Answering}

\IEEEtitleabstractindextext{%
\begin{abstract}
    Despite its importance for assessing the effectiveness of communicating information visually, fine-grained recallability of information visualisations has not been studied quantitatively so far.
In this work, we propose a question-answering paradigm to study visualisation recallability and present VisRecall --- a novel dataset consisting of 200 visualisations that are annotated with crowd-sourced human (N~=~305) recallability scores obtained from 1,000 questions of five question types.
Furthermore, we present the first computational method to predict recallability of different visualisation elements, such as the title or specific data values. 
We report detailed analyses of our method on VisRecall and demonstrate that it outperforms several baselines in overall recallability and FE-, F-, RV-, and U-question recallability.
Our work makes fundamental contributions towards a new generation of methods to assist designers in optimising visualisations.

\end{abstract}

}

\maketitle

\IEEEdisplaynontitleabstractindextext

\IEEEpeerreviewmaketitle

\IEEEraisesectionheading{\section{Introduction}\label{sec:introduction}}

Memorability is an intrinsic, global, and stimulus-driven perceptual property that is important for better comprehension of visual stimuli~\cite{bainbridge2017memorability, bylinskii2021memorability}.
A growing body of work has studied image recognisability -- one of the most fundamental attributes of memorability, both from a perceptual~\cite{standing1973learning, bainbridge2017memorability} and a computational~\cite{isola2011makes, khosla2015understanding} perspective.
Recognisability has also been studied in information visualisations and previous work has revealed specific attributes that make visualisations memorable~\cite{borkin2013makes}.
Recognisability measures whether a visualisation looks familiar or novel~\cite{standing1973learning}.
A visualisation that has unique features may stand out more and may therefore be more memorable. 
However, recognisability does not capture how effective a visualisation is in conveying information to observers.
Other works have therefore studied \textit{recallability} -- a concept that goes beyond memorability~\cite{borkin2015beyond} by quantifying \textit{what} viewers remember from a visualisation~\cite{rust2020understanding}. 
Despite its importance and potential for designing better information visualisations, a deeper understanding of which characteristics of visualisations influence recallability, and in which way, is currently missing.

Current methods to assess recallability rely on visualisation experts to assign a qualitative score to self-reported free-text descriptions of viewers~\cite{borkin2015beyond}.
This approach is cumbersome and only provides a single score representing overall recallability while hiding the contribution of individual visualisation characteristics.
While \citet{borkin2015beyond} noted the importance of titles for recallability on visualisations, \citet{polatsek2018exploring} conducted three low-level analytical tasks, focusing on visual elements with extrema, or specific values.
These works inspired us to quantify visualisations' recallability by looking into specific types of visualisation elements, such as the title, elements with extrema, or distinct data points.

To quantify recallability, we propose to adopt a question-answering paradigm, similar to visual question answering (VQA)~\cite{antol2015vqa} that has become widely popular in computer vision. 
While originally introduced for natural images~\cite{antol2015vqa, goyal2017making}, a VQA \cmr{dataset} was also \cmr{proposed} for information visualisations~\cite{kahou2017figureqa}.
\cmr{In our work, instead of proposing computational models of reasoning and correctly answering questions about images}, we evaluate the performance of human observers in answering questions about visualisations and use their performance as a subjective measure of information visualisation recallability.

In this work, to quantify fine-grained recallability of information visualisations, we design and execute a question-answering based study to collect \datasetNameShort: a novel visualisation dataset with 200 visualisations, which contains 1,000 high-quality questions annotated by visualisation experts and crowd-sourced human recallability scores.
Our work is inspired by and extends prior task taxonomy on visualisations~\cite{amar2005low, polatsek2018exploring} to define fine-grained recallability scores through five question types: identifying the title or theme, finding extrema, filtering data elements, retrieving values, and understanding structure~(\autoref{ss:dataset_questiontype}).
Through our analyses of \datasetNameShort, we make several interesting findings:
the highest recallability across question types occurs in questions that are about the title or the general theme~(T-question), which is significantly higher than other question types.
Moreover, a 10-second encoding duration is sufficient for most visualisation types, including \textit{bar, pie, line,} and \textit{scatter plots}.
Based on \datasetNameShort, we further present \methodNameShort, a novel method based on convolutional neural networks (CNNs) to predict one overall and five fine-grained recallability scores, one for each question type.

Our contribution is threefold: (1) We adapt a question-answering paradigm to quantify \cmr{overall and} fine-grained recallability of information visualisations.
(2) We collect \datasetNameShort, a novel visualisation dataset with human recallability scores (N\,=\,305) from 1,000 questions and five question types.
(3) We propose a computational model that predicts fine-grained recallability of visualisations. %
As such, our work points the way towards new methods and tools to create more effective information visualisations. 

\section{Related Work}\label{sec:relatedwork}
Our work is related to previous works on 1) image memorability, 2) perception and memorability of visualisations, and 3) \cmr{chart question answering (chart QA)} datasets.

\subsection{Image Memorability}
A pioneering study~\cite{standing1973learning} reported a strong capability of humans to recognise what they have seen before even up to 10,000 images, which is denoted as ``image recognition memory''.
\citet{isola2011makes,isola2013makes} have demonstrated that memorability is an observer-independent property, which only depends on images~\cite{brady2008visual, isola2011understanding}.
Furthermore, previous studies have proven that memorability could be reliably quantified for individual images by asking subjects to report whether images are novel or familiar.
Large-scale memorability datasets have been collected for natural images, such as SUN-Mem~\cite{isola2011makes}, Figrim~\cite{mancas2013memorability} and LaMem~\cite{khosla2015understanding}.
With the rise of deep learning, deep convolutional neural networks were proposed as computational methods to predict image memorability~\cite{khosla2015understanding, perera2019image, jaegle2019population}.
Recent work also integrated visual attention into the memorability prediction model~\cite{fajtl2018amnet}.
Meanwhile, recallability is a complementary memory task to visual recognition~\cite{haist1992relationship}, which requires subjects to view images and then recall what they have seen~\cite{yonelinas2002nature}.
One previous work found that sketch-based methodologies can improve the recall of a sampling distribution from an experiment~\cite{hullman2017imagining}.
Several recent studies are consistent with the conclusion that image memorability variation may be distinct for recognition and recall tasks~\cite{bainbridge2019drawings, rust2020understanding}.
Based on this, our work is the first to improve understanding of recallability characteristics and the factors that influence it on information visualisations.

\subsection{Perception and Memorability of Visualisations}
 
Pioneering works in the visualisation community have examined how different data types and tasks influence human perception~\cite{cleveland1984graphical, kosslyn1989understanding, pinker1990theory}.
\citet{inbar2007minimalism} reported that people prefer over-embellishment (i.e., ``chart junk'') instead of Tufte's minimalist design~\cite{tufte1985visual}.
\citet{bateman2010useful} further claimed that the ``chart junk'' improves recognisability but is not essential for understanding the visualisation. This triggered a series of studies evaluating the impact of style on memorability and comprehensibility~\cite{borgo2012empirical, moere2012evaluating, shu2020makes}.
The effect of specific factors or components on recall memory has been investigated, such as interaction~\cite{kim2012does}, prior knowledge~\cite{kim2017explaining}, title~\cite{borkin2015beyond, kong2019trust} and text redundancy~\cite{borkin2015beyond}.
\citet{borkin2013makes} studied visualisation memorability on the MASSVIS dataset, and their follow-up work~\cite{borkin2015beyond} further conducted online crowd-sourcing studies to quantify both recognisability and recallability.
However, there are two main drawbacks to the previous recallability quantification procedure. Firstly, the method used to recall quality annotations is subjective and cumbersome. 
In addition, visualisation experts are necessary to attribute these scores.
Secondly, the description quality score scale with only four possible values is too coarse to represent a visualisation. %
To overcome these limitations, we introduce question answering as a powerful paradigm to quantify the recallability of information visualisations.
Through multiple questions and answers on different visualisation characteristics, we propose a novel computational model to predict not only overall but also fine-grained recallability based on five different question types. 

\subsection{Chart Question Answering~(Chart QA) \cmr{Datasets}}
Despite the importance of information visualisations, chart QA datasets have only been proposed in recent years.
FigureQA~\cite{kahou2017figureqa} was the first chart QA dataset. 
Images were plotted in simple and fully synthesised visualisations in five visualisation classes, along with polar questions.
DVQA~\cite{kafle2018dvqa} focused specifically on the problem of visual reasoning on bar charts, which was used as a corpus for generating the topic of chart QA.
PlotQA~\cite{methani2020plotqa} and LEAF-QA~\cite{chaudhry2020leaf} synthesised their question-answer pairs based on crowd-sourced question templates from real-world data sources to increase variety.
\cmr{One follow-up work collected human answers to the DVQA dataset~\cite{kafle2020answering} as baselines, which inspired us to use the human performance as a subjective measure of information visualisation recallability.}

As a conclusion, the question-answering setting has not yet been used for memorability studies on visualisations, and current chart QA datasets are synthesised from simple templates with limited content, making it a distance away from real-world visualisations.
However, chart QA provides an interesting means to quantify recallability.
In our work, we evaluate and obtain recallability scores by asking users questions and validating their answers. 
Therefore, we present the design of our novel adaptation of a question-answering-based study on information visualisations and our novel VisRecall dataset in the next section.

\section{\datasetNameShort Dataset}\label{sec:dataset}
\begin{figure*}[t]
    \centering
    \includegraphics[width=\textwidth]{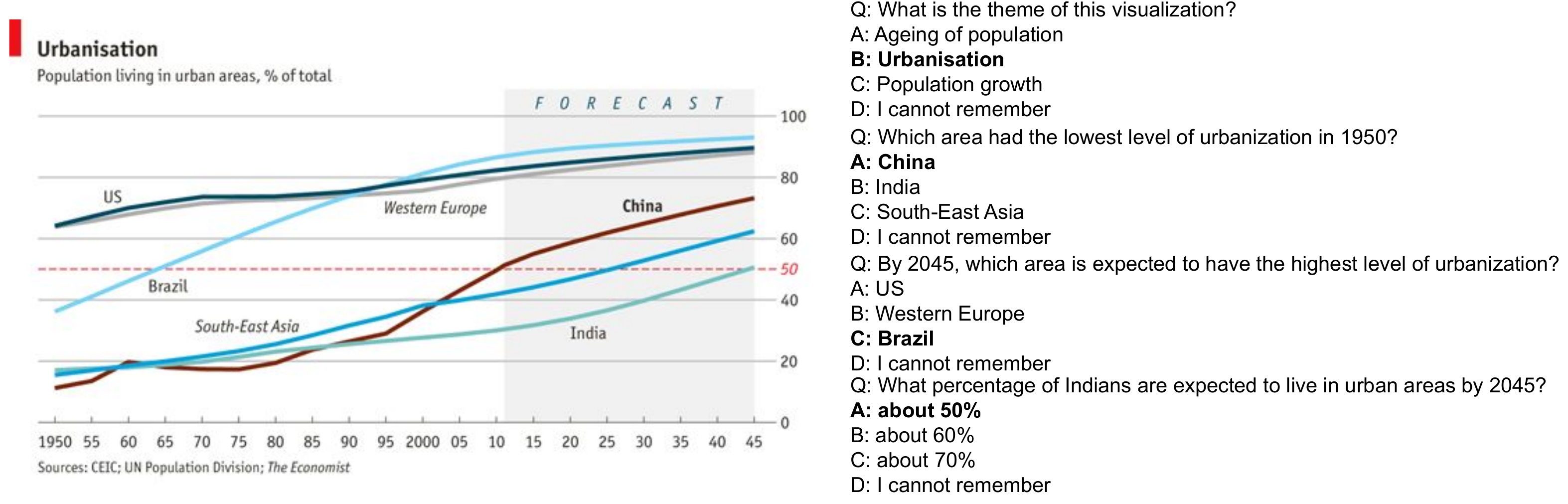}
    \caption{Sample visualisation with multiple-choice questions from \datasetNameShort. Five types of questions were designed by experts, which are questions regarding the title~(T-questions), understanding structure or trend~(U-questions), finding extrema~(FE-questions), filtering elements~(F-questions) and retrieving values~(RV-questions). Each figure has at least two question types. The correct answer to each question is shown in \textbf{bold}. Image sourced from MASSVIS~\protect\cite{borkin2013makes}.
    }
    \label{fig:dataset_example}
\end{figure*}

The currently available recallability scores on the visualisation dataset MASSVIS~\cite{borkin2013makes,borkin2015beyond} are annotated from free-text descriptions.
However, its procedure to quantify recallability is coarse and cumbersome.
Meanwhile, visual question-answering~(VQA) datasets~\cite{antol2015vqa} selectively target elements of visualisations in different question-answer pairs, making it a suitable setting to quantify memorability objectively and efficiently. 
Under the question-answering paradigm, different tasks can be represented as different types of questions to viewers, and consequently, recallability is quantified by the accuracy in answering those questions.

Towards quantifying recallability, we propose the \datasetName~(\datasetNameShort) --- a dataset consisting of 200 real-world information visualisations with crowd-sourced human recallability scores~(N\,=\,305) obtained from 1,000 questions in five question types~(see Figure~\ref{fig:dataset_example}). Visualisations in our dataset are mainly sourced from the MASSVIS dataset~\cite{borkin2013makes} to enable better alignment with prior works on this topic.
The recognisability scores are also collected to replicate the previous memorability studies~\cite{borkin2013makes,borkin2015beyond}. Our dataset and code are accessible at the link\footnote{ \url{https://doi.org/10.18419/darus-2826}}.

\subsection{Visualisation Collection and Question Types}
\label{ss:dataset_questiontype}
We randomly selected a subset of 200 visualisations from the MASSVIS dataset~\cite{borkin2013makes}.
Notably, we excluded all infographics from our collection, since infographics have the highest recognisability and recallability compared to all other types of visualisations~\cite{borkin2015beyond}.
However, scatter plots represented only 5\,\% of the sampled subset.
Therefore, we collected 20 additional scatter plot visualisations by crawling the web through search engines~(Google, Bing) using the keyword ``scatter plots''. Then, we replaced some bar plots with the web-crawled plots to balance the visualisation type classes.
The final distribution of visualisation types is: 56 bar plots, 45 line plots, 27 scatter plots, 22 pie plots, 25 tables and 25 \textit{others}. Those visualisations that don't belong to any of the first five types are categorised as \textit{others}, including box charts, isotype charts, and other complex visualisations.

\cmr{
When creating the questions for our VisRecall dataset, we had to identify question types that are not only suitable for static information visualisations, since our images are primarily sourced from the MASSVIS dataset but also applicable to all visualisations.
Therefore, inspired by prior work, we selected five question types: T-questions, U-questions, FE-questions, F-questions, and RV-questions.
T-questions were questions regarding the title or the visualisation theme, which were used to analyse how the existence of the title influenced the description scores of visualisation recallability~\cite{borkin2015beyond}.
U-questions were about understanding the plot structure~\cite{methani2020plotqa} or the general trend~\cite{chaudhry2020leaf}.
For the remaining three question types, we followed the work by Polatsek et al.~\cite{polatsek2018exploring} who picked these question types for the visualisation types from MASSVIS.
The three question types were FE-questions (finding an extremum attribute value), F-questions~(filtering visualisation elements based on specific criteria) and RV-questions~(retrieving values for a specific visualisation element).
}

All question-answering data were created by five data visualisation experts.
They were asked to provide five questions per visualisation, and every visualisation has at least two question types.
Each question corresponds to four possible answer options.
Only one option is correct, two other options are choices with similar, yet incorrect answers, and the last option is always ``I cannot remember''. See supplementary material for question examples.
All annotations were saved separately in standard JSON files for each visualisation. There are 193, 150, 178, 99, and 64 visualisations in \datasetNameShort that have at least one T-, FE-, F-, RV-, and U-question, respectively.

\begin{figure*}[t]
    \centering
    \includegraphics[width=0.9\textwidth]{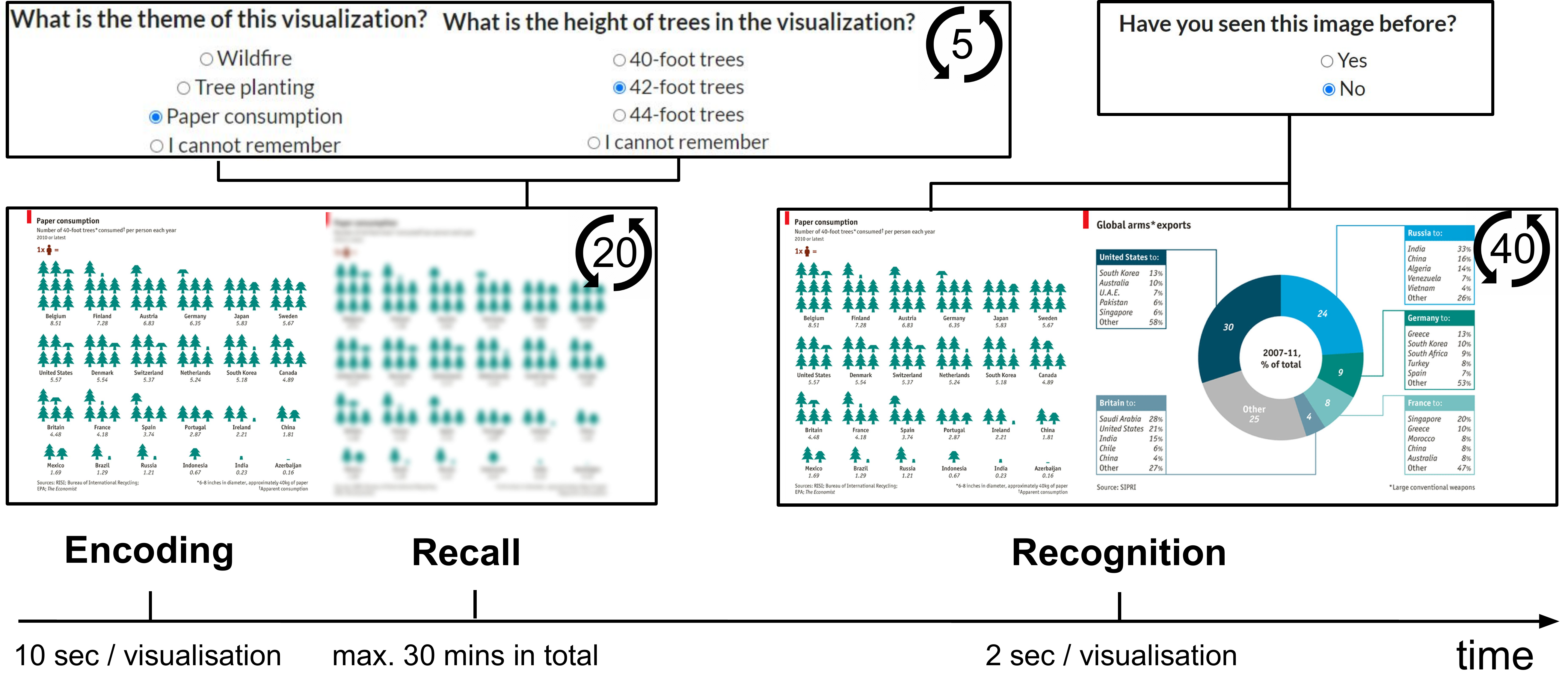}
    \caption{Experiment design. From left to right: Visualisations are shown to viewers for a fixed duration in the ``Encoding'' phase. In the ``Recall'' phase, visualisations are blurred and each has a multiple-choice question next to it with a single correct answer.
    Finally, visualisations are shown to viewers for 2 seconds in the ``Recognition'' phase. The numbers in the circular arrows indicate the number of repetitions.
    }
    \label{fig:dataset_studydesign}
\end{figure*}

\textbf{T-question.}
T-questions are about the title or the general theme of the plot and do not require any reasoning.
Example questions: \textit{What is the title of the visualisation?}, \textit{What is the theme of the visualisation?}
For the incorrect choices in T-questions, we either replaced keywords or phrases with words of similar, but different meanings, such as changing \textit{car thefts} to \textit{car accidents} or \textit{car manufacturers}, or using titles from other visualisations, such as using \textit{Expectations On House Prices Above 2009 Projections} and \textit{HIV Prevalence in Women Aged 15-49 Years by Region, 1990-2007} as incorrect choices for \textit{Covered Transactions by Sector and Year, 2009-2011}.

\textbf{FE-question.}
These are questions about finding extreme values in the visualisation that fulfil certain conditions, without asking any exact numbers.
Example questions: \textit{Which area had the lowest level of urbanization in 1950?}, and \textit{Which particle is the latest discovered?}
We used other elements that appeared in the visualisation as incorrect answer choices.
As seen in Figure~\ref{fig:dataset_example}, \textit{India} and \textit{South-East Asia} were the incorrect alternative choices for \textit{China} in the question \textit{Which area had the lowest level of urbanization in 1950?}.

\textbf{F-question.}
These are questions about filtering data elements based on specific criteria.
Example questions: \textit{Which particle is Bosons?} and \textit{What is the source of the data?}
For F-questions, we either changed keywords to their synonyms, or used other elements that appeared in visualisations as incorrect alternative choices, such as using \textit{Electron} and \textit{Muon} for \textit{Photon} in the question \textit{Which particle is Bosons?}.

\textbf{RV-question.}
These are questions about retrieving a specific value located in the plot. %
Example questions: \textit{What is the maximum percentage of aid allocated?} and \textit{What percentage of Indians are expected to live in urban areas by 2045?}~(see Figure~\ref{fig:dataset_example}). Example incorrect choices: \textit{about 60\,\%} and \textit{about 70\,\%} for \textit{What percentage of Indians are expected to live in urban areas by 2045?}, and the correct answer is \textit{about 50\,\%}.
 
\textbf{U-question.}
These are questions about understanding the structure or the trend of a visualisation.
Example questions: \textit{What does the purple curve represent?} and \textit{What decreases as time goes by?} Example incorrect choices: for structure questions, other elements appearing in the visualisation are used, such as using \textit{Red} and \textit{Blue} as incorrect choices for \textit{Green} in the question \textit{What color stands for Residents?} As for questions about understanding trends, the choices are \textit{increasing}, \textit{decreasing} and \textit{almost the same}.

\subsection{Crowd-sourcing Study Set-up \& Participants}
Our study design is illustrated in Figure~\ref{fig:dataset_studydesign}.
In the encoding phase of our study, study participants were shown a sequence of visualisations for a fixed duration. We followed the 10-second encoding phase in a prior memorability study~\cite{borkin2013makes}, and also conducted the study with a 20-second encoding phase to see the impact of encoding duration on recallability.
We asked participants to memorise as much of the information presented in each visualisation as possible. To advance from the encoding phase to the recall phase, our study participants had to click on the ``next'' button.
In the recall phase, each visualisation was shown at 50\% of the size from the encoding phase and blurred by a 24-pixel Gaussian filter to make the text unreadable. 
The question orders were predefined to avoid the situation where some questions might provide answers to other questions. The blurry visualisation was shown with a single multiple-choice question. The presentation order of the first three multiple-choice options for each question was randomly shuffled once and fixed for all participants, while the option \textit{I cannot remember} always appeared last. The following question would be shown only if the participant clicked the next button, and they could not return to the previous question. This setting was to avoid providing hints in upcoming questions.
\cmr{Before running our user study, we did preliminary tests with three designs: showing visualisations one-by-one, two-by-two, and four-by-four. 
In the one-by-one setting, we found the task too easy with very high recallability scores.
The four-by-four task, i.e. first encoding four different visualisations before the recall phase, was found too difficult. 
Therefore, as a trade-off and to reduce the effect of working memory~\cite{owen2005n}, we empirically selected the two-by-two setting for our user study.}
In each set, the encoding phase of two images were presented, followed by their recall phase, before repeating the process for the next set of two images.
Then, the recognition phase involved an online memorability game similar to prior work~\cite{borkin2013makes}. Study participants were presented with a sequence of images, and they had to select if they had seen this visualisation before.
In each Human Intelligence Task~(HIT), 40 blurred images were shown for 2 seconds each. The images in the recognition phase contained 20 visualisations that were the same in the recall phase, and 20 fillers from a different group.
Finally, participants were asked to provide anonymous feedback on the study design in a questionnaire.

To support the study, we implemented the procedures in a web application. We then integrated our application into an existing crowd-sourcing toolbox that worked well with the Amazon Mechanical Turk~(MTurk) platform~\cite{2020TurkEyes}.
We deployed our experiment on MTurk to collect recallability and recognisability scores on all 200 visualisations, splitting them randomly into ten groups of 20 visualisations per HIT. Visualisation types were balanced among all groups~(see \cmr{Figure 1 in} supplementary material). MTurk workers could participate in multiple HITs. To participate in one of our HITs, a worker had to be a Master Worker approved by MTurk as a quality check. 
Master Workers are top workers rated by MTurk who have consistently demonstrated high quality work. 
Workers were paid \$\,4.00 for completing each HIT. To ensure data quality, we filtered out 467 HITs (305 workers) if the answers were all ``Yes'' or ``No'' in the recognition task. For each visualisation, we received an average of 40.4 ($\sigma$\,=\,16.9) valid responses.
All workers were distributed in various educational levels: 8.2\,\% two-year degree, 56.9\,\% four-year degree, 22.3\,\% master’s degree or higher, and 12.6\,\% other\,/\,unreported. The age groups were 44.1\,\% in 25\,-\,34, 28.5\,\% in 35\,-\,44, 12.4\,\% in 45\,-\,54 and 9.9\,\% over 55.
In the anonymous feedback form at the end of our study, most workers responded positively, with two examples being: ``Great self test for capable of memory power'' and ``After taking survey, I'm really getting interested in learning data plots and visualisations''.

\subsection{Data Analysis}

\textbf{Recallability Formulation.}
For each question, we measured the recall accuracy as follows:
$Acc=\frac{RA}{RA+WA},$
where \textit{RA} is the number of correct answers, and \textit{WA} is the number of wrong answers, including the number of \textit{I cannot remember} answers.
If we focus on viewers who have selected choices excluding \textit{I cannot remember}, the accuracy can be computed as:
$Acc'=\frac{RA}{RA+WA-CNR}$, where CNR stands for the number of \textit{I cannot remember}.
Averaging all questions of type \textit{t} in a visualisation gives us the recallability by question type and is computed as:
$Rec_{t}=\frac{1}{n}\sum_{i=1}^{n}Acc(i), question_i \in t.$
By averaging all questions in a visualisation, we have the overall recallability of a visualisation as: 
$Rec=\frac{1}{n}\sum_{i=1}^{n}Acc(i)$.

\begin{figure}
    \centering
    \includegraphics[width=\linewidth]{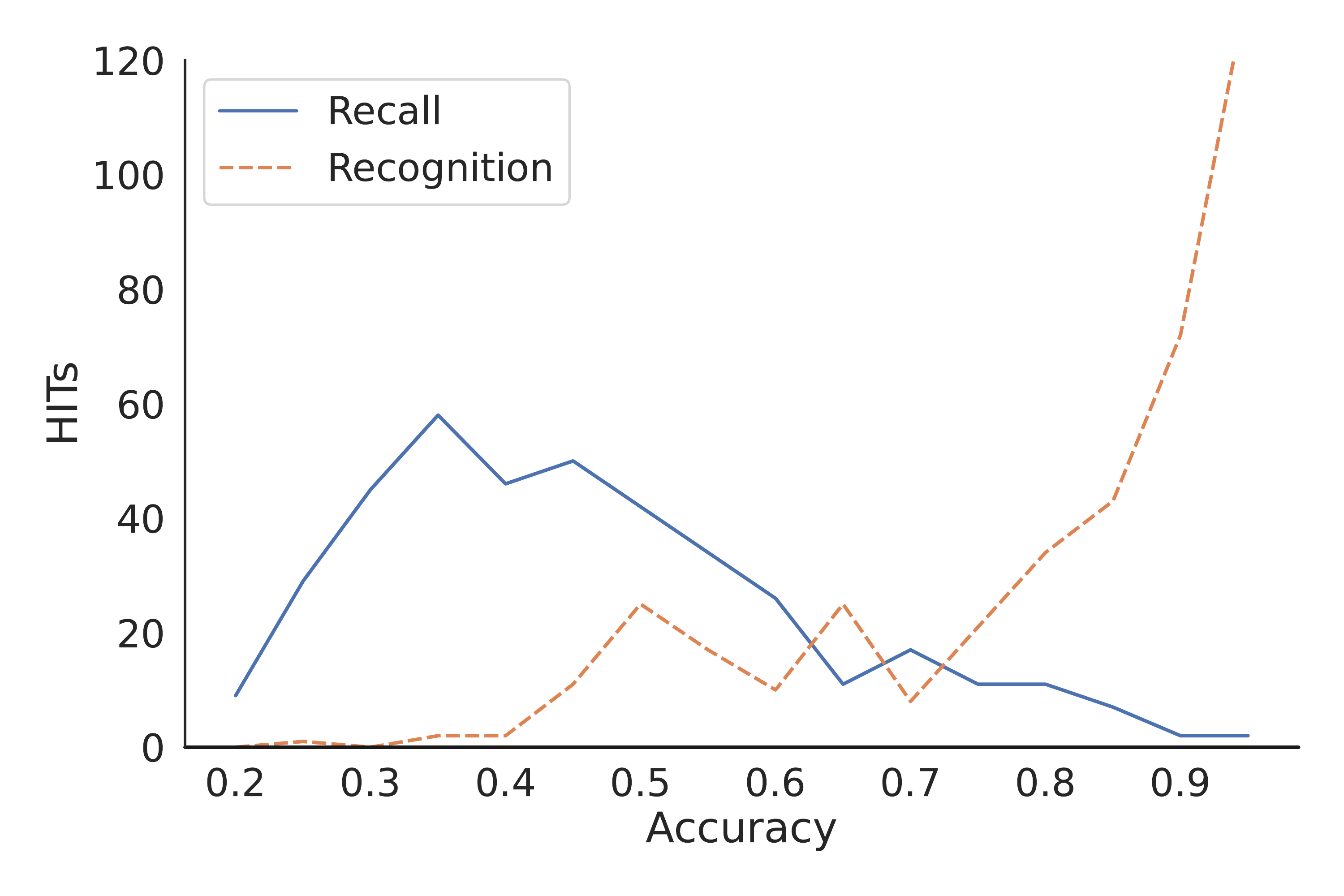}
    \caption{Recall and Recognition accuracy over all 404 HITs. Participants can recognise most of the visualisations easily, but only answer \cmr{less than} half of the questions correctly.}
    \label{fig:dataset_acc}
\end{figure}

\textbf{HIT-wise Recallability.}
HIT-wise recallability as well as recognition accuracy across HITs~(N\,=\,404) are shown in Figure~\ref{fig:dataset_acc}. 63.9\,\% of HITs have a recognition accuracy higher than 0.85, and 34.83\,\% are higher than 0.95, which shows that our study participants could easily recognise most of the visualisations ($\mu$\,=\,0.83). Meanwhile, they could only answer \cmr{less than} half of the questions correctly~($\mu$\,=\,0.49).

\begin{figure}[t]
    \centering
    \includegraphics[width=\linewidth]{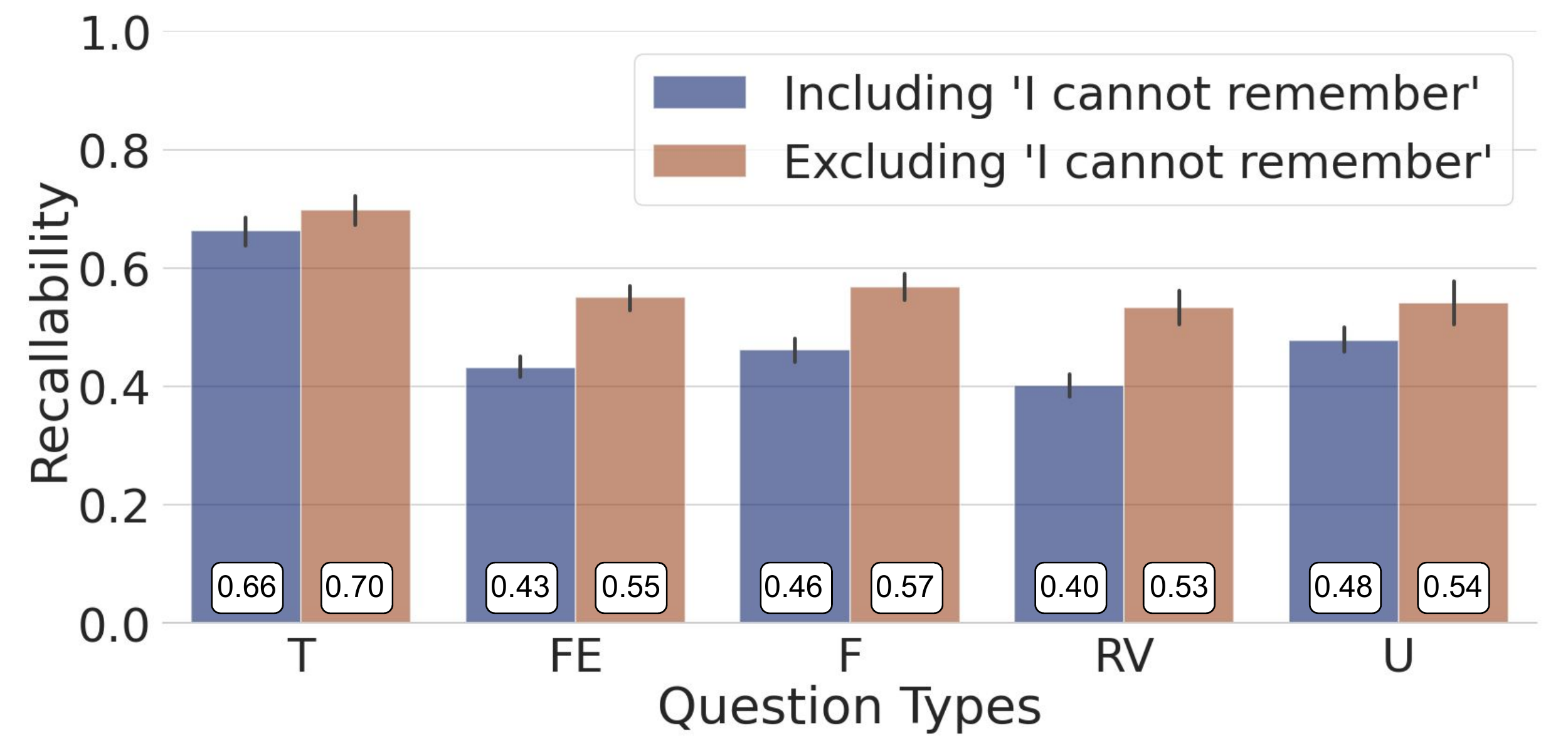}
    \caption{Recallability scores by question type. T-questions have significantly higher recallability scores \cmr{than} all other question types~(FE-, F-, RV-, and U-questions). Additionally, 24.7\,\% of the viewers selected \textit{I cannot remember} in RV-questions, \cmr{while} only 5.1\,\% of the viewers selected \textit{I cannot remember} in T-questions.}
    \label{fig:dataset_type_acc}
\end{figure}

\textbf{Fine-grained Recallability by Question Type.}
Figure~\ref{fig:dataset_type_acc} illustrates that T-questions have the highest recall accuracy among all question types both when including \textit{I cannot remember}~($\mu$\,=\,0.66), and excluding \textit{I cannot remember} ($\mu$\,=\,0.69). The accuracy of T-questions is significantly higher than other question types~(t\,(1969)\,=\,18.87, p\,<\,0.001). 24.7\,\% of viewers selected \textit{I cannot remember} in RV-questions, and 21.4\,\%, 18.8\,\%, 11.7\,\% for FE-, F- and U-questions, respectively. Only 5.1\,\% of the study participants selected \textit{I cannot remember} in T-questions.
We observed a mean proportion of 19.1\,\% ($\sigma$\,=\,13.0\,\%) of study participants who selected \textit{I cannot remember} in all visualisations. The lowest proportion is 3\,\%, while more than 50\,\% of participants selected \textit{I cannot remember} in seven visualisations. Figure~\ref{fig:dataset_drate} shows visualisations with the most and least \textit{I cannot remember} answers from \datasetNameShort.
We observe that a high visualisation complexity is common among those visualisations with the most \textit{I cannot remember} answers.

\begin{figure}[t]
    \centering
    \includegraphics[width=\linewidth]{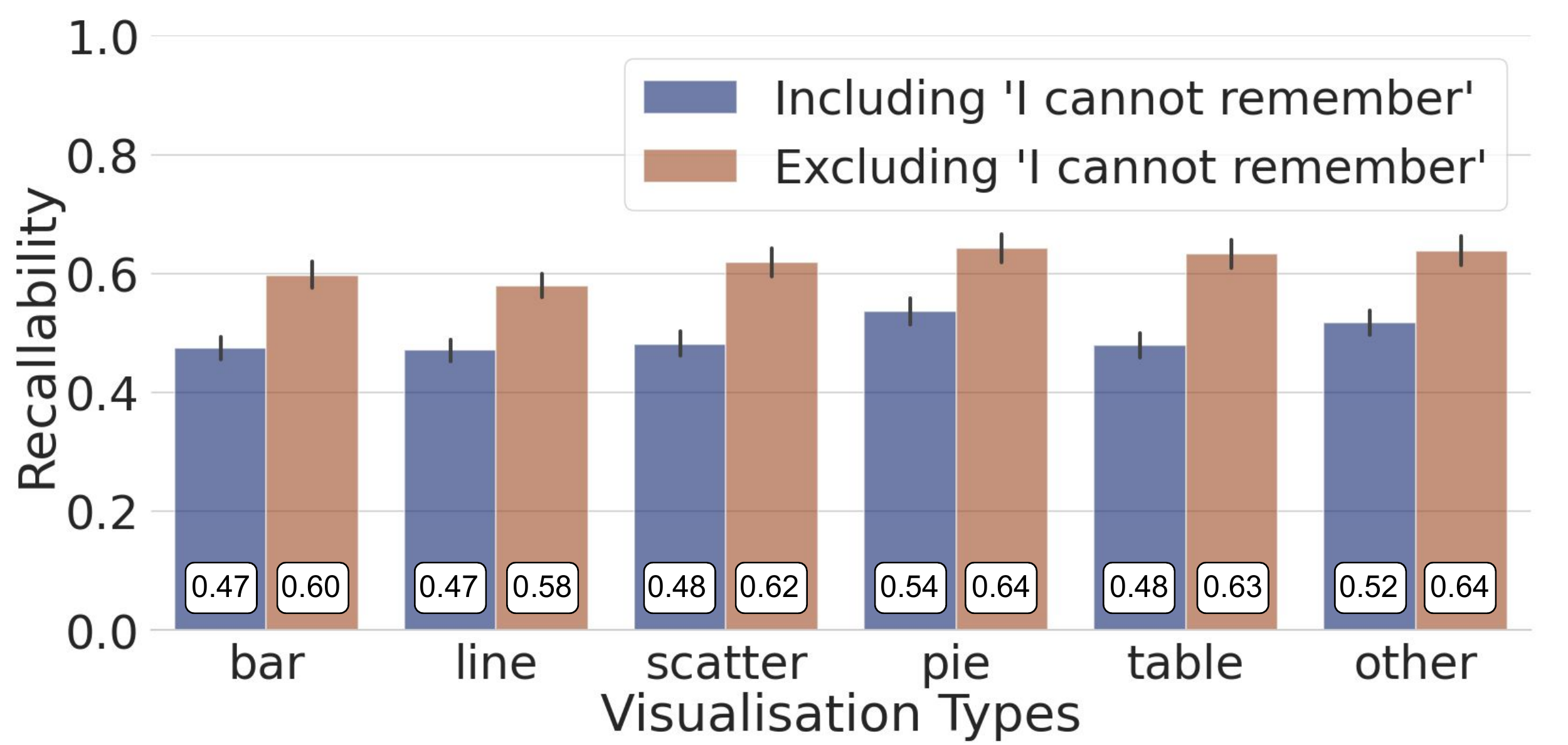}
    \caption{Recallability scores by visualisation type. Pie plots have significantly higher recallability scores compared with all other visualisation types~(bar, line, scatter plots, tables, and others).}
    \label{fig:dataset_vistype_acc}
\end{figure}

\textbf{Fine-grained Recallability by Visualisation Type.}
Figure~\ref{fig:dataset_vistype_acc} illustrates the recallability scores by visualisation type. A one-way ANOVA test is applied across visualisation types, and we observed a significant difference for both excluding \textit{I cannot remember}~(F\,=\,4.412, p\,<\,0.001), and including \textit{I cannot remember}~(F\,=\,6.916, p\,<\,0.001). Post-hoc analyses \cmr{with} Tukey's HSD~\cite{tukey1949comparing} confirmed that the recallability scores of pie plots are significantly higher than any other visualisation types including \textit{I cannot remember}~(for all pairs, p\,<\,0.001). For excluding \textit{I cannot remember}, line plots are significantly lower than pie plots~(t\,=\,3.725, p\,=\,0.003) and \textit{others}~(t\,=\,3.458, p\,=\,0.007).

\begin{figure}[t]
    \centering
    \includegraphics[width=\linewidth]{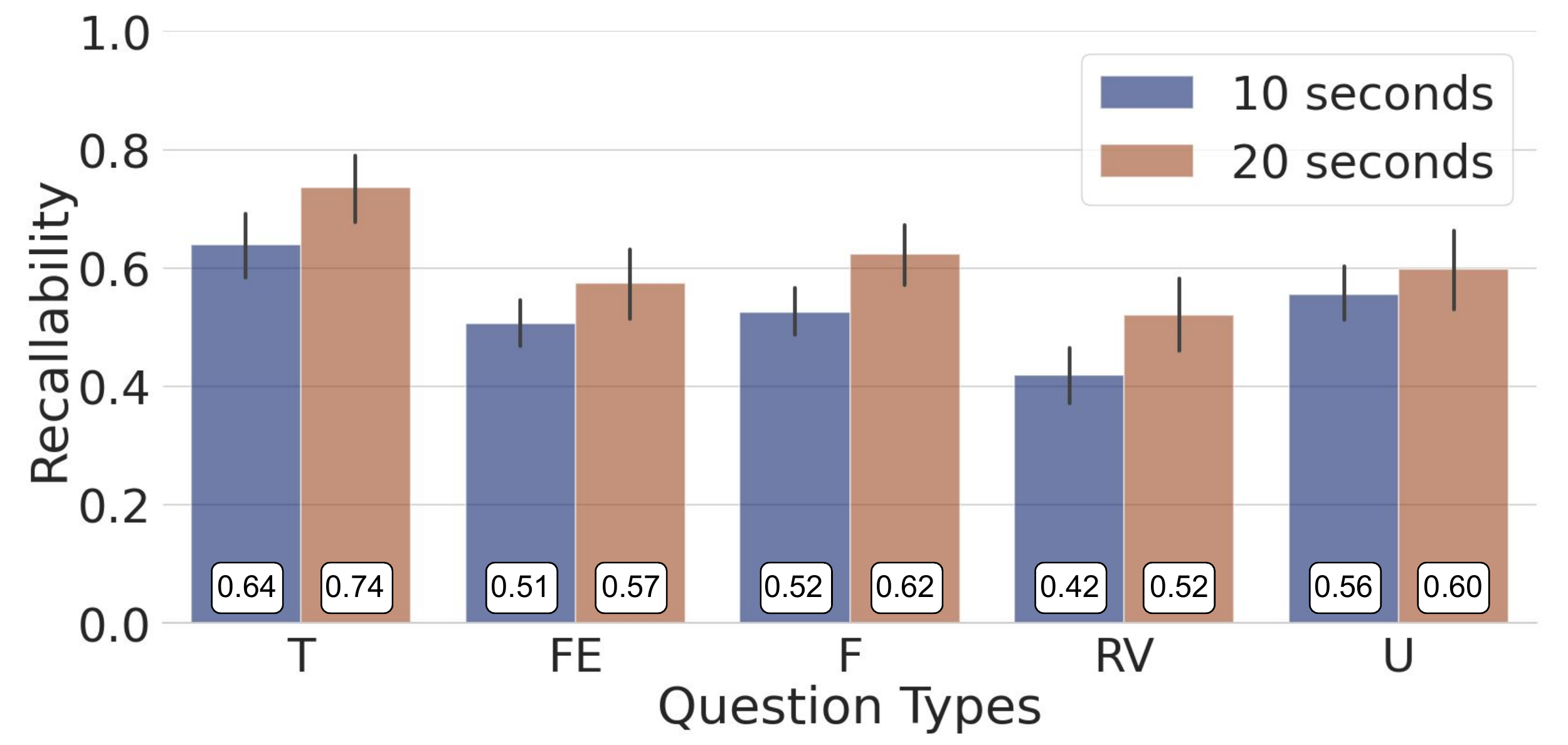}
    \caption{Recallability scores under a 10-second and 20-second encoding phase by question type in one MTurk group.}
    \label{fig:dataset_20s_qtype}
\end{figure}

\begin{figure}[t]
    \centering
    \includegraphics[width=\linewidth]{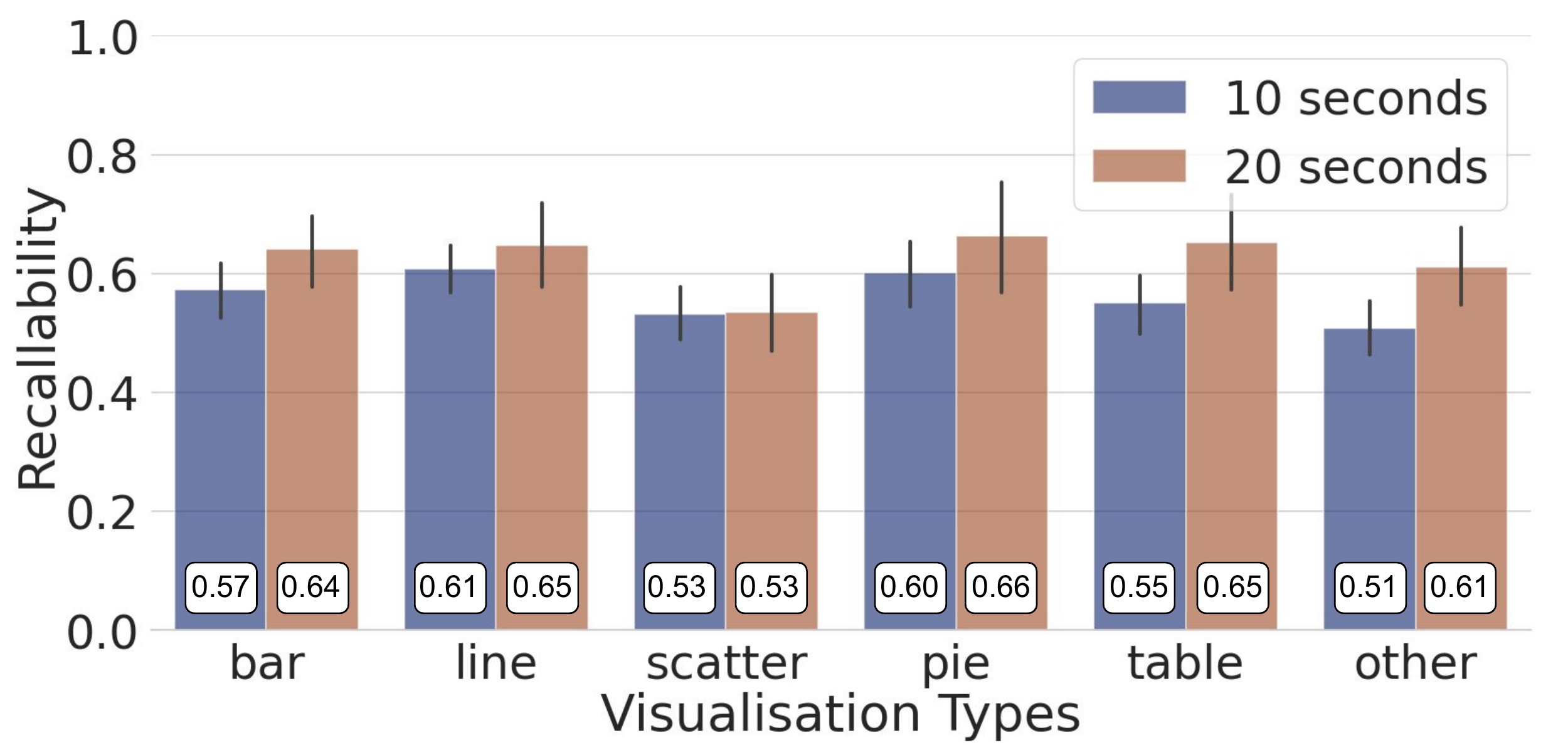}
    \caption{Recallability scores under a 10-second and 20-second encoding phase by visualisation type in the same MTurk group as Figure~\ref{fig:dataset_20s_vistype}.}
    \label{fig:dataset_20s_vistype}
\end{figure}

\begin{figure*}[t]
    \centering
    \includegraphics[width=\linewidth]{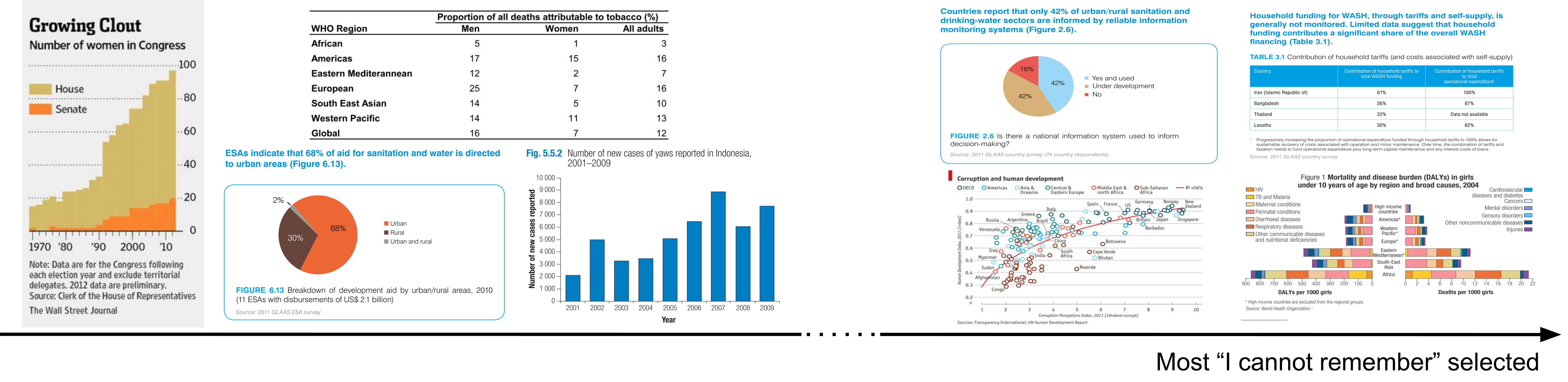}
    \caption{Example visualisations with the most and fewest answers \textit{I cannot remember} from \datasetNameShort. We observed a higher degree of visualisation complexity for those with multiple \textit{I cannot remember} answers.}
    \label{fig:dataset_drate}
\end{figure*}

\textbf{Encoding Duration.}
The study of the 20-second encoding phase was conducted in two randomly selected MTurk groups.
We observed significant improvement of recallability in one group~(t\,(198)\,=\,2.284, p\,=\,0.023) from 10-second ($\mu$\,=\,0.51, $\sigma$\,=\,0.51) to 20-second encoding phase ($\mu$\,=\,0.57, $\sigma$\,=\,0.22), but not in the other group~(t\,(198)\,=\,1.627, p\,=\,0.105).
The recallability scores by question type is shown in Figure~\ref{fig:dataset_20s_qtype}. Prolonging the 10-second encoding phase to 20 seconds, the \cmr{recallability} scores of each question type all increased.
\cmr{No significant improvements of recallability scores were found in T-questions (t\,(59)\,=\,1.367, p\,>\,0.05), F-questions (t\,(59)\,=\,1.796, p\,>\,0.05), FE-questions (t\,(59)\,=\,1.474, p\,>\,0.05), RV-questions (t\,(59)\,=\,1.951, p\,>\,0.05), or U-questions (t\,(59)\,=\,0.830, p\,>\,0.05).}
Figure~\ref{fig:dataset_20s_vistype} \cmr{illustrates} the recallability scores by visualisation type.
Significant improvements of recallability scores are found in tables~(t\,(59)\,=\,2.144, p\,=\,0.036) and \textit{others}~(t\,(59)\,=\,2.969, p\,=\,0.009), but not in pie plots~(t\,(59)\,=\,1.141, \cmr{p\,>\,0.05}), bar plots~(t\,(59)\,=\,1.675, \cmr{p\,>\,0.05}), line plots~(t\,(59)\,=\,1.052, \cmr{p\,>\,0.05}), or scatter plots~(t\,(59)\,=\,0.817, \cmr{p\,>\,0.05}).

\begin{figure*}[t]
    \centering
    \includegraphics[width=0.42\linewidth]{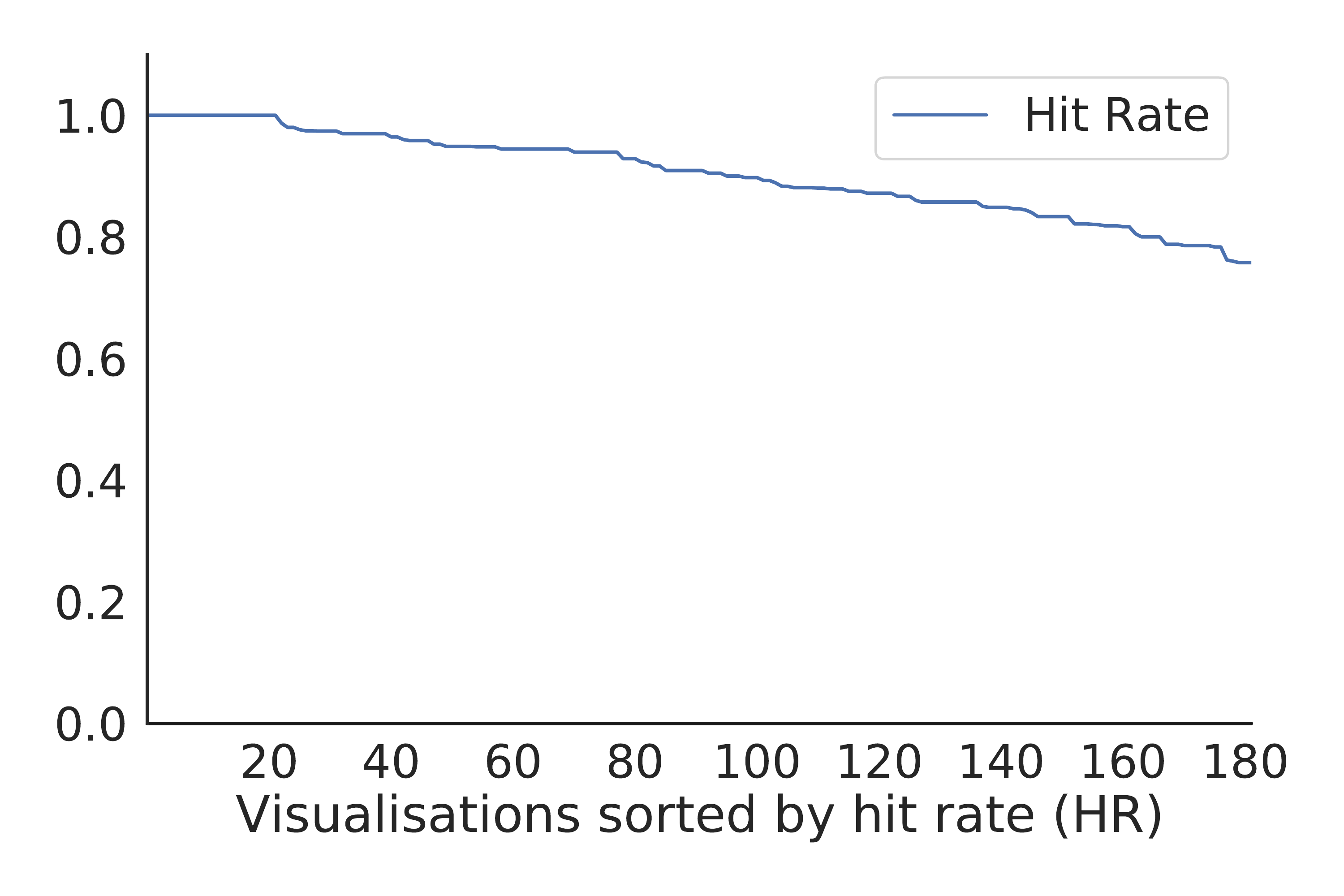}
    \includegraphics[width=0.57\linewidth]{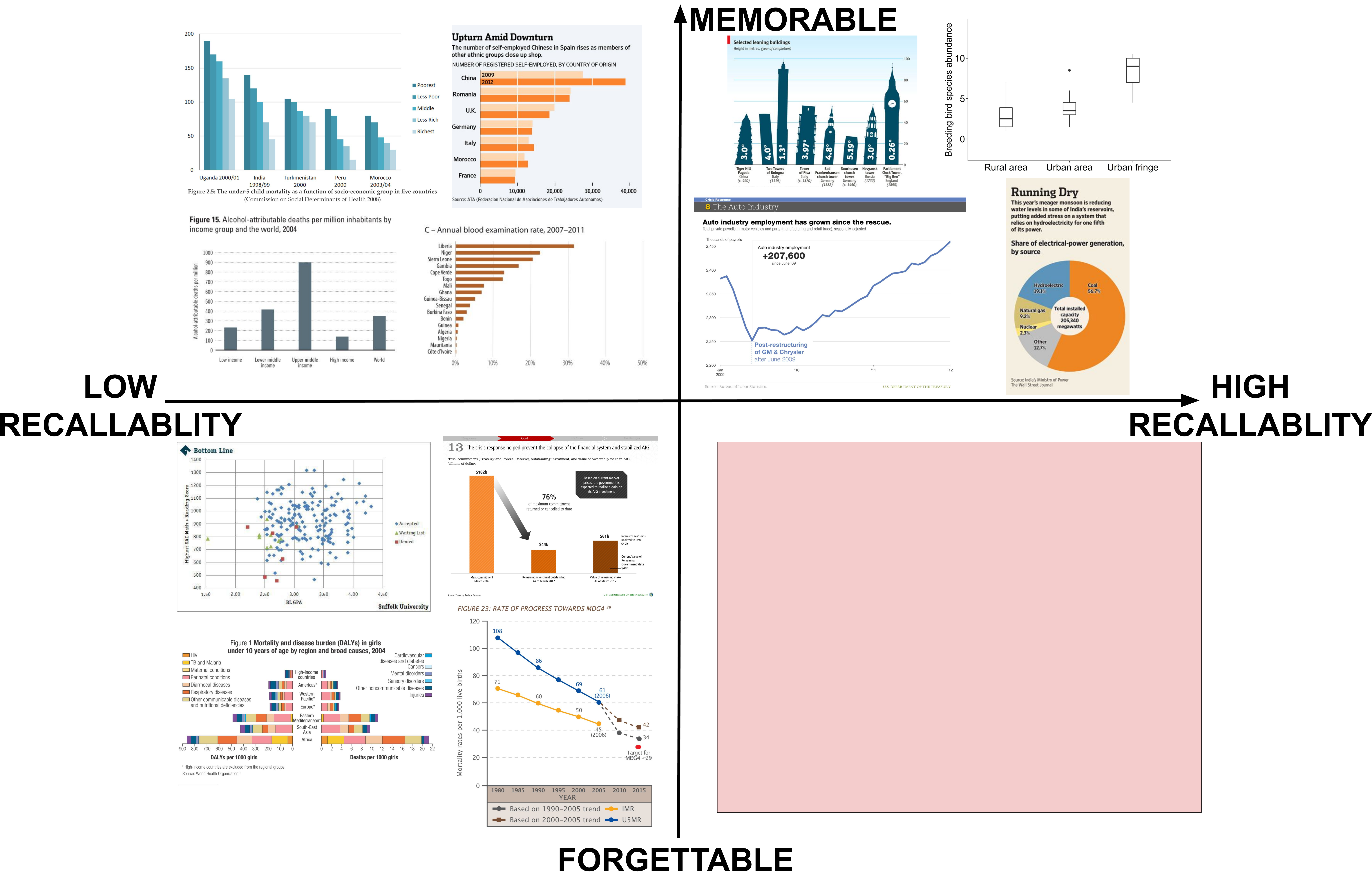}
    \caption{Left: Raw hit rate~(HR) of target visualisations from the recognition phase. Right: The highest and lowest ranked visualisations~(within 15\,\%) across recognisability~(memorability) and recallability in \datasetNameShort. The x-axis represents the recallability score computed from overall visualisation question accuracy (independent of question type), and the y-axis represents the memorability score from previous work~\cite{borkin2015beyond}.
    }
    \label{fig:discussion_memrecall}
\end{figure*}

\textbf{Recallability and Recognisability: a Comparison to Prior Work.}
To the best of our knowledge, the description quality in~\cite{borkin2015beyond} is the closest work to ours on the quantification of recallability, where free text descriptions of what participants recall about the visualisations were recorded. Description quality was rated from 0 to 3 where 0 was a completely incorrect description, and 3 was a precise description regarding the topic and at least one detail~\cite{borkin2015beyond}. %
We found 31 visualisations with description quality in our VisRecall dataset, and calculated the average description quality scores and the mean visualisation accuracy of all questions in the overlapped visualisations. The Pearson's correlated coefficient~(CC) between description quality scores and mean visualisation accuracy is 0.36 with \textit{I cannot remember}, while it is 0.35 without \textit{I cannot remember}.

For a comparison to prior work on recognisability~\cite{borkin2013makes, borkin2015beyond}, we also calculated the memorability (or recognisability) score on \datasetNameShort.
According to~\citet{borkin2013makes}, the hit rate~(HR) and false alarm rate~(FAR) were computed as:
$HR=\frac{HITS}{HITS+MISSES}$ and
$FAR=\frac{FA}{FA+CR}$.
Then, the recognisability~(memorability) of a visualisation was measured as:
$d'=Z(HR)-Z(FAR)$,
where Z was the inverse cumulative Gaussian distribution.
Figure~\ref{fig:discussion_memrecall}\,(left) shows the distribution of the raw HR scores of all visualisations from the recognition phase. 
Figure~\ref{fig:discussion_memrecall}\,(right) shows the highest and lowest ranked visualisations across recognisability~(memorability) and recallability from our \datasetNameShort dataset.
\cmr{The full memorability~(recognisability) and recallability scores of all visualisations are available in supplementary material}.
Visualisations in each quadrant were ranked highest or lowest 15\,\% among all visualisations.

\textbf{Data-ink ratio.}
Data-ink ratio is a commonly used visual attribute introduced by~\citet{tufte1990envisioning}. A high data-ink ratio visualisation contains a large share of ink presenting information about data.
\cmr{Following the previous annotation procedure in \cite{borkin2015beyond},} three visualisation researchers independently rated the data-ink ratio for each visualisation in the VisRecall dataset. \cmr{Data-ink ratio was rated from 1 to 3 where 1 was a low propotion of ink that was related to data, and 3 was high.} The ranking was directly applied if more than two researchers agreed. In cases when all three researchers gave different rankings, the visualisation was reviewed and discussed by all three researchers for a consensus.
We observed that the high data-ink ratio group has the highest recallability score ($\mu$\,=\,0.621), compared with $\mu$\,=\,0.599 for the middle data-ink ratio, and $\mu$\,=\,0.606 for the low data-ink ratio.
\cmr{A} one-way ANOVA test was applied between data-ink ratio groups. Still, no significance was observed in either excluding \textit{I cannot remember}~(F\,=\,1.134, p\,=\,0.322) or including \textit{I cannot remember}~(F\,=\,2.43, p\,=\,0.088).

\textbf{Visual Density.}
Visual density is another visual attribute to rate the overall density of visual elements without distinguishing between data and non-data elements~\cite{borkin2013makes}.
The annotation \cmr{procedure} of visual density was the same as data-ink ratio.
\cmr{Visual density was rated from 1 to 3 where 1 was low density of visual elements in the visualisation, and 3 was high.}
We observed that the high visual density group has the highest recallability score ($\mu$\,=\,0.619), compared with $\mu$\,=\,0.608 for middle visual density, and $\mu$\,=\,0.600 for low visual density.
A one-way ANOVA test is applied across visual density groups, but no significance is observed in either excluding \textit{I cannot remember}~(F\,=\,0.740, p\,=\,0.478) or including \textit{I cannot remember}~(F\,=\,0.245, p\,=\,0.782).

\section{Computational model for predicting fine-grained Recallability}\label{sec:method}
Our analyses on \datasetNameShort yielded several insights on recallability in information visualisations.
There are currently no baseline methods, for predicting \cmr{either} overall recallability or fine-grained recallability.
Existing computational models only aimed at predicting memorability, also known as recognisability~\cite{khosla2015understanding, fajtl2018amnet}.
Therefore, we propose \methodNameLong~(\methodNameShort), a lightweight and effective neural network for recallability prediction.

\subsection{Model Architecture}

\begin{figure*}[htbp]
    \centering
    \includegraphics[width=0.9\linewidth]{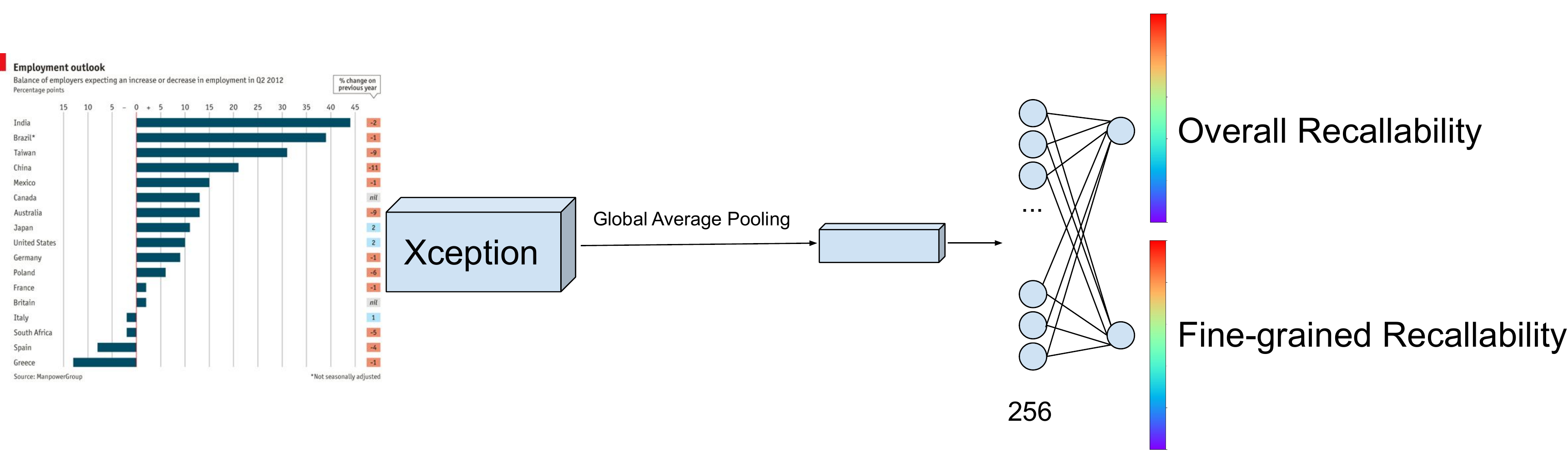}
    \caption{Method overview. \methodNameShort leverages the Xception model~\cite{chollet2017xception} to effectively encode spatial information. Then, a global average pooling layer, a dense layer with 256 neurons, and finally a dense layer with 2 neurons are sequentially connected. One output neuron predicts the general recallability score, and the other one predicts the fine-grained recallability score.}
    \label{fig:Network}
\end{figure*}

We extend and build on state-of-the-art architectures from other computer vision tasks, such as semantic segmentation~\cite{chollet2017xception, chen2017deeplab} and image classification~\cite{Simonyan15, he2016deep}, and use such methods as the backbone of our architecture.
We design our \methodNameShort with the specific goal of predicting both overall and fine-grained recallability scores in one single model~(see~\autoref{fig:Network} for an overview).
Inspired by UMSI~\cite{fosco2020predicting}, the current state-of-the-art architecture for visual importance prediction on graphic designs, we employ the Xception~\cite{chollet2017xception} model to effectively encode spatial information. Then, a global average pooling layer, a dense layer with 256 neurons, and finally a dense layer with 2 neurons are sequentially connected. One output neuron predicts the general recallability score, and the other one predicts the fine-grained recallability score.

\subsection{Implementation Details \& Model Training}
We trained \methodNameShort using weights obtained from the Xception model -- which was pretrained on ImageNet~\cite{deng2009imagenet}.
\methodNameShort was trained with the Adam~\cite{kingma2014adam} optimizer with a learning rate of 0.002 and 1:1 Mean Squared Error~(MSE) joint loss for the two branches predicting the overall recallability score and the fine-grained recallability score.
We averaged all five questions for each image to prepare the ground truth of overall recallability scores.
To train our \methodNameShort to predict fine-grained recallability scores for a certain question type, we only used those visualisations that contained that question type from \datasetNameShort. There are 193, 150, 178, 99, and 64 visualisations with at least one T-, FE-, F-, RV-, and U-question, respectively. Five-fold cross-validation was applied to all evaluation processes.
All experiments were conducted on a single NVIDIA 2060 Super GPU with 8GB VRAM.

\textbf{Baseline Methods.} \xspace
Since no previous computational models focused on predicting recallability on visualisations, we designed three methods as baselines. We replaced the Xception feature encoder in \methodNameShort with VGG-16~\cite{Simonyan15} and ResNet-34~\cite{he2016deep} as the two baselines. %
We trained all baseline models for 10 epochs on \datasetNameShort starting from ImageNet~\cite{deng2009imagenet} pretrained weights. 
We used the Adam optimizer~\cite{kingma2014adam} with a learning rate of 0.002 and MSE loss for training.

\subsection{Model Evaluation}

The prediction error is calculated as the mean squared error between the human and the predicted recallability scores.
We compared the prediction error of our \methodNameShort method to the two baselines
VGG-16 and ResNet-34. %
Table \ref{table:experiment_quantitative} summarises fine-grained recallability prediction error on \datasetNameShort under a 5-fold cross-validation evaluation.
We used the MSE to evaluate the prediction error.
Results showed that \methodNameShort outperformed the baselines under overall recallability and four fine-grained recallability scores, with a MSE of 0.035 for overall recallability, and 0.021, 0.022, 0.017, 0.043 for FE-, F-, RV-, and U-questions respectively. 
ResNet-34 was the best performing method for T-questions with a MSE of 0.047, while our \methodNameShort was second with a MSE of 0.052.

\begin{table*}[t]
    \centering
    \caption{Prediction error~(MSE) of fine-grained recallability on \datasetNameShort under 5-fold cross-validation evaluation. Best results are shown in \textbf{bold}, second-best are \underline{underlined}.}
    \begin{tabular}{ccccccc}
\toprule
    \textbf{Methods} & \textbf{Overall} & \textbf{T} & \textbf{FE} & \textbf{F} & \textbf{RV} & \textbf{U} \\ %
\midrule
    \methodNameShort~(ours) & $\bm{0.035\pm0.005}$ & $\underline{0.052\pm0.009}$ & $\bm{0.021\pm0.003}$ & $\bm{0.022\pm0.004}$ & $\bm{0.017\pm0.004}$ & $\bm{0.043\pm0.025}$ \\%& 25.6\\
    ResNet-34~\cite{he2016deep} & $0.043\pm0.013$ & $\bm{0.047\pm0.015}$ & $0.068\pm0.024$ & $0.070\pm0.042$ & $\underline{0.043\pm0.008}$ & $\underline{0.050\pm0.018}$ \\ %
    VGG-16~\cite{Simonyan15} & $\underline{0.036\pm0.013}$ & $0.053\pm0.017$ & $\underline{0.054\pm0.019}$ & $0.076\pm0.029$ & $0.057\pm0.010$ & $0.059\pm0.025$ \\ %
\bottomrule
    \end{tabular}
    \label{table:experiment_quantitative}
\end{table*}

\textbf{Ablation study.}\xspace
We further carried out an ablation study to investigate how each fine-grained recallability score influences overall recallability (see Table \ref{table:experiment_abl_overall}). 
\cmr{With} RecallNet, the overall recallability trained with T-questions has the lowest \cmr{MSE} of 0.030 and the most stable variance of 0.006. \cmr{With} ResNet-34~\cite{he2016deep}, the overall recallability trained with RV-questions has the lowest \cmr{MSE} of 0.029 and the most stable variance of 0.008. \cmr{With} VGG-16, the overall recallability trained with T-questions has the lowest \cmr{MSE} of 0.037 and the most stable variance of 0.007.

\begin{table*}[t]
    \centering
    \caption{Ablation study on the prediction error~(MSE) of how fine-grained recallability influences the overall recallability. Best results in each row are shown in \textbf{bold}.}
    \begin{tabular}{cccccc}
\toprule
    \textbf{Methods} & \textbf{T} & \textbf{FE} & \textbf{F} & \textbf{RV} & \textbf{U} \\
\midrule
    \methodNameShort (ours) & $\bm{0.030\pm0.006}$ & $0.079\pm0.052$ & $0.032\pm0.008$ & $0.035\pm0.013$ & $0.172\pm0.215$ \\
    ResNet-34~\cite{he2016deep} & $0.043\pm0.013$ & $0.078\pm0.087$ & $0.060\pm0.035$ & $\bm{0.029\pm0.008}$ & $0.033\pm0.013$ \\
    VGG-16~\cite{Simonyan15} & $\bm{0.037\pm0.007}$ & $0.046\pm0.022$ & $0.041\pm0.019$ & $0.079\pm0.053$ & $0.077\pm0.011$ \\
\bottomrule
    \end{tabular}
    \label{table:experiment_abl_overall}
\end{table*}

\section{Discussion}\label{sec:discussion}
First, we underline the novelty of \datasetNameShort and its potential in applications such as chart QA~\cite{kahou2017figureqa}. Second, we discuss how recallability and recognisability are different yet connected. Then, several interesting insights from our analyses are reported. Finally, the limitations and future work are discussed.

\textbf{\datasetNameShort Dataset.}
\datasetNameShort is the first dataset to introduce fine-grained recallability on an information visualisation dataset as well as high-quality \cmr{question-answering} annotations.
The recallability scores are metrics that reveal human performance with a specific type of question.
With rich annotations of the elements necessary for the answers, the recallability score of a certain question could be converted into 2D spatial representations~(e.g. recallability heatmaps).
The recallability maps could be introduced as an additional feature input to downstream tasks, such as chart QA.
Additionally, \datasetNameShort is a novel visualisation question-answering dataset that uses real-world, visually rich visualisations coming in part from the MASSVIS dataset.
The questions for chart QA datasets~\cite{methani2020plotqa,chaudhry2020leaf} were collected by regular crowd workers. 
In contrast, all the questions in our \datasetNameShort came from visualisation experts, which promises a higher quality of questions than chart QA datasets.
Moreover, most visualisations in chart QA datasets~\cite{methani2020plotqa} are generated automatically. However, when it comes to real-world visualisations, the layout information is usually missing, and researchers have to retrieve it, often by manual annotation~\cite{borkin2015beyond}, which is time-consuming and constrains the dataset size.
The introduction of recallability to the question-answering setting and the high quality of visualisations and questions enable \datasetNameShort to trigger fundamental studies in chart QA.

\textbf{Recallability vs. Recognisability~(Memorability).}
The bottom-right quadrant in Figure~\ref{fig:discussion_memrecall}~(right) is completely empty, which means that there are no such visualisations with high recallability~(top 15\,\%) and low memorability~(bottom 15\,\%) in \datasetNameShort. 
This means that \textit{visualisations have to be sufficiently memorable before they become recallable}.
The visualisations in the top-right quadrant share some characteristics, like a big and highlighted title and some explanatory text.
Meanwhile, the visualisations in the top-left quadrant of Figure~\ref{fig:discussion_memrecall}~(right) have high recognisability and low recallability.
Compared to the top-right quadrant, visualisations in the top-left quadrant are less recallable. 
All visualisations in the top-left quadrant are simple monotone plots with few embellishment~(e.g. isotype plots).
The visualisations in the bottom-left quadrant are easily forgettable and hard to recall. 
These visualisations are usually overly complex and don't have meaningful titles or additional explanatory text to convey key messages.
Compared to the bottom-left quadrant, all the visualisations in the top-left and top-right quadrant are always with titles, which aligns well with the findings in previous studies~\cite{borkin2013makes, borkin2015beyond}.
However, the recallability between the data-ink ratio and visual density groups is not significantly different. Either those visual features are not highly correlated with recallability, or the size of our \datasetNameShort prevented the confirmation of significance.
\cmr{Nevertheless}, our study on \datasetNameShort validated previous results and provided interesting insights into how recallability and recognisability~(memorability) are different yet connected.

\textbf{Free Recall vs. Question-Answering-cued Recall.}
The low correlated relationship~(CC\,=\,0.35) between description quality~\cite{borkin2015beyond} and our recallability score drew our attention.
The description quality generated from prior work was from free-text descriptions without any context, but our recallability was computed from the mean accuracy of five multiple-choice questions per image.
The low correlated relationship suggests that the information~(context) in multiple-choice questions might be an essential factor that influenced recallability.
One possible explanation is that our study provided cues for visualisations which mitigated the memory decaying process~(forgetting)~\cite{jonides2008mind}.

\textbf{Impact of Encoding Duration on \cmr{Recallability}.}
The analysis on encoding duration provided several insights~(see Figures~\ref{fig:dataset_20s_qtype} and \ref{fig:dataset_20s_vistype}).
Those text-heavy and complex visualisations (tables, \textit{others}) are more sensitive to viewing duration, and a 10-second encoding phase is sufficient for most basic visualisation types~(pie, bar, line, and scatter plots).
Filtering data and retrieving value questions (F- and RV-questions) improved more than the other three question types~(T-, FE-, and U-questions). It suggests that a prolonged encoding phase is more beneficial to those questions that require detailed answers~(F- and RV-questions).

\textbf{Limitations.}
There is always a trade-off between quality and quantity, which was also the case when designing and collecting our \datasetNameShort dataset.
Due to the increasing workload in designing high-quality questions for the question-answering settings specifically targeted for each visualisation, the scale of \datasetNameShort became relatively small.
We conducted a preliminary evaluation using Grad-CAM~\cite{selvaraju2017grad}, which is a method used for understanding and explaining the predictive behaviour of CNN-based models.
However, our qualitative analysis did not reveal any generalisable patterns that can be directly linked to higher-level visualisation features such as visual density or data-ink ratio.
To allow more explainable models for recallability prediction, it is essential to extend the size of \datasetNameShort.

\textbf{Future Work.}
How recallability can be applied to reality is a fundamental question. Visualisation type recommendation is one practical use case for a visualisation recommendation system~\cite{hu2019vizml}.
Prior research has proposed ways to decide whether line or scatter plots are more suitable for time series data~\cite{wang2017line}. One possible application is to make use of recallability scores to recommend a visualisation type for given data. For visualisations that demonstrate the same data but with different visualisation types, our RecallNet might be useful in recommending a visualisation type that maximises recallability.

In the future, we plan to enrich \datasetNameShort with more complex visualisation \cmr{types} such as box, radar and combination plots.
Furthermore, gaze behaviour analysis in a question-answering setting on information visualisations has not yet been studied.
However, it is a fundamental step to understand the human visual attention system while viewing visualisations.
While physical laboratory studies require special-purpose eye tracking equipment, online crowd-sourcing studies or gaze estimation from substitution devices~(e.g., mouse, web camera) can be used as a proxy to human attention. 
In the future, we will investigate such methods to collect human attention data and extend \datasetNameShort with such annotations.

\section{Conclusion}
This work presented a novel adaptation of a question-answering-based study to collect \datasetNameShort, a novel visualisation dataset with 200 \cmr{real-world} visualisations annotated with crowd-sourced human recallability scores in five question types, along with a deep convolutional network to predict fine-grained recallability of visualisations.
Overall, this work made a substantial leap towards quantifying fine-grained recallability scores on information visualisations and envisions several potential applications, \cmr{such as visualisation optimisation}.

\ifCLASSOPTIONcompsoc
  \section*{Acknowledgments}
\else
  \section*{Acknowledgment}
\fi

Y. Wang was funded by the Deutsche Forschungsgemeinschaft~(DFG, German Research Foundation)~-~Project-ID 251654672~-~TRR~161.
M. B\^{a}ce was funded by a Swiss National Science Foundation (SNSF) Early Postdoc. Mobility Fellowship (grant number 199991).
A. Bulling was funded by the European Research Council (ERC; grant agreement 801708).

We would like to thank Sruthi Radhakrishnan for the creation of visualisations for our application, Maurice Koch, Guanhua Zhang, Adnen Abdessaied, and Florian Strohm for data annotation, Dominike Thomas for paper editing support, as well as Yue Jiang, Daniel Weiskopf and Nils Rodrigues for helpful comments on this paper.

\ifCLASSOPTIONcaptionsoff
  \newpage
\fi

\bibliographystyle{IEEEtranN}
\bibliography{main}

\begin{IEEEbiography}[{\includegraphics[width=1in,height=1.25in,clip,keepaspectratio]{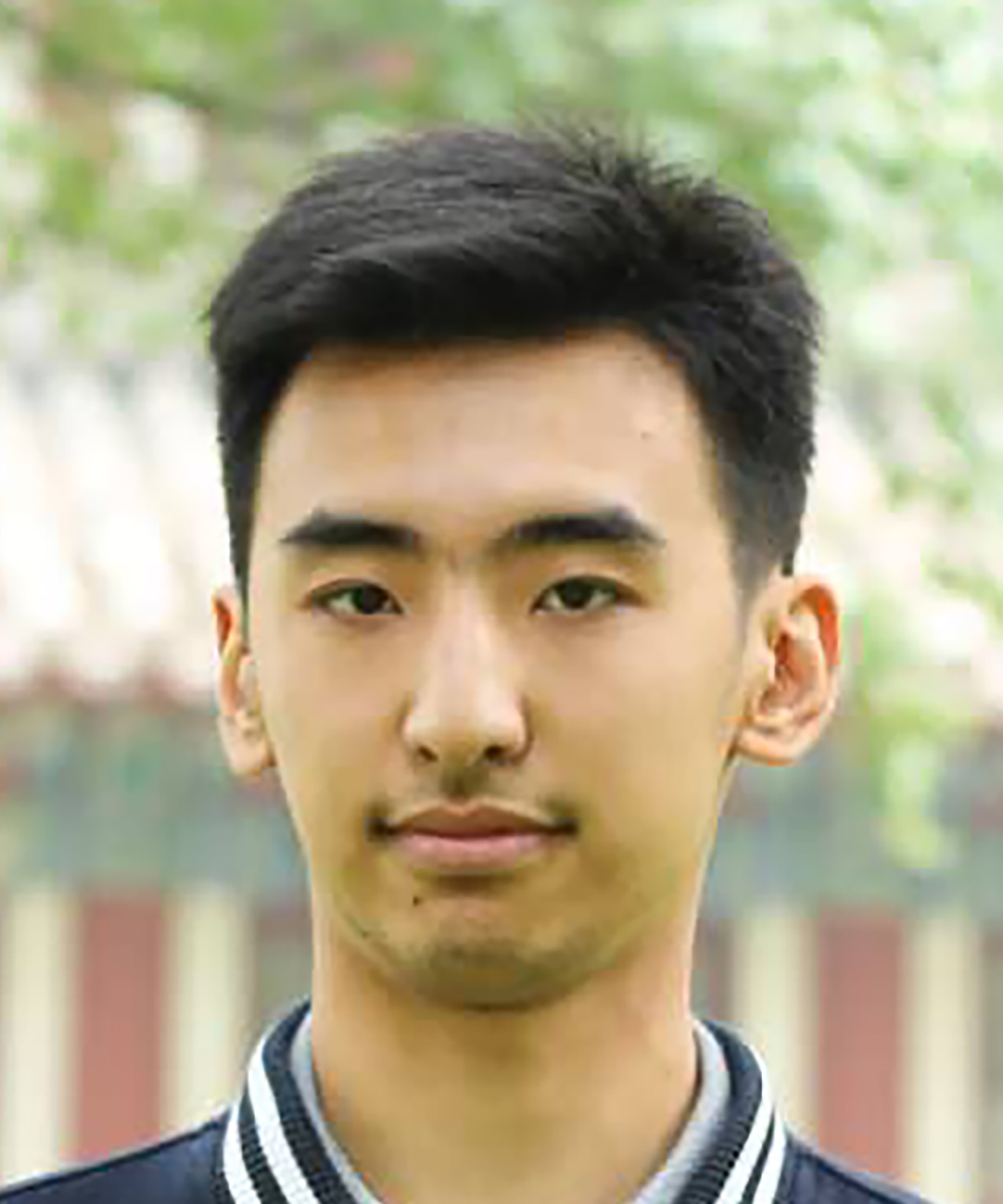}}]{Yao Wang}
is a PhD student at the University of Stuttgart, Germany. He received his BSc. in Intelligence Science and Technology and MSc. in Computer Software and Theory both from Peking University, China, in 2017 and 2020, respectively. His research interests include computer vision and human-computer interaction, with a focus on visual attention modelling on information visualizations.
\end{IEEEbiography}

\begin{IEEEbiography}[{\includegraphics[width=1in,height=1.25in,clip,keepaspectratio]{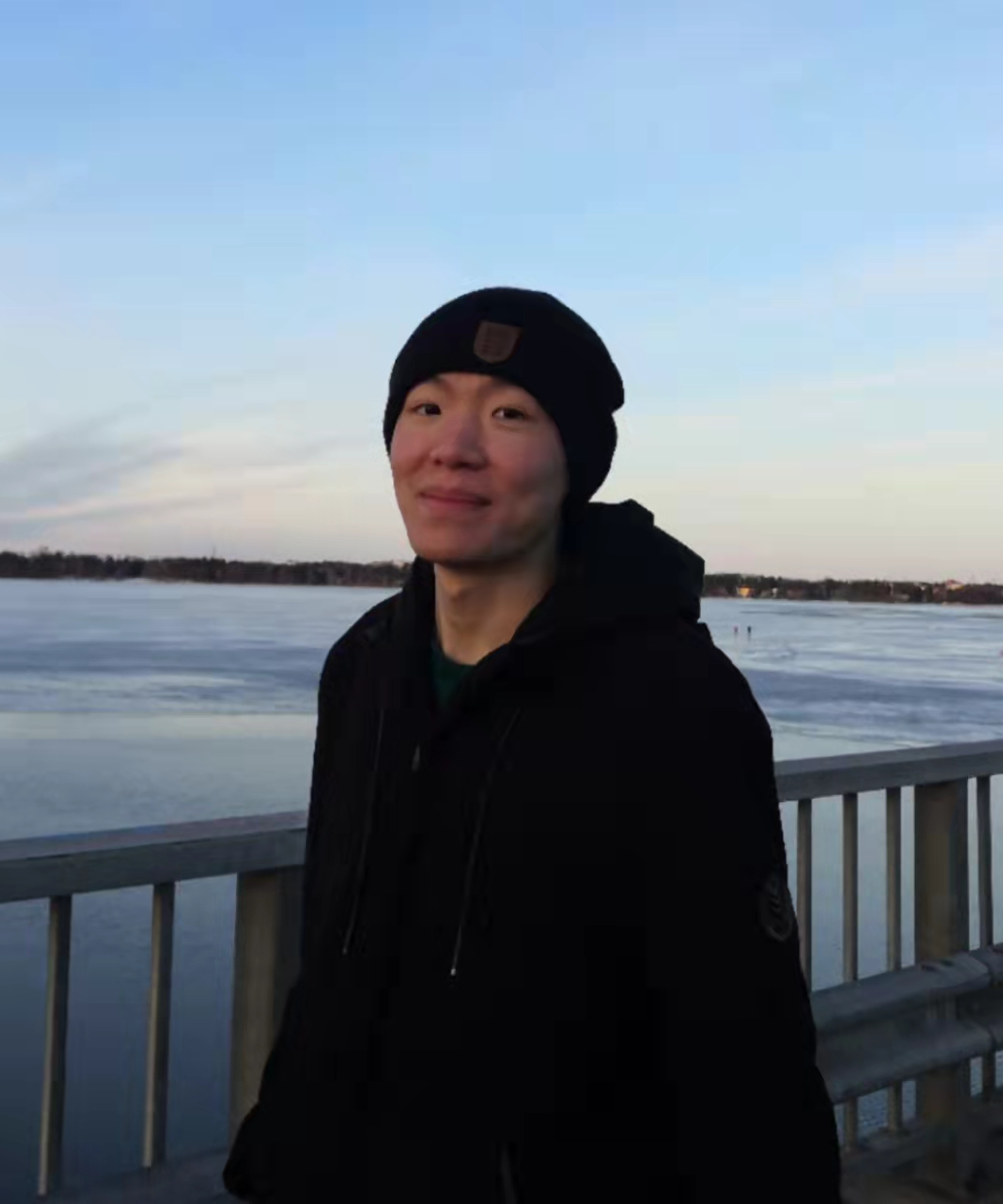}}]{Chuhan Jiao}
is an MSc. student in Computer Science at Aalto University. He received his BEng. in Computer Science and Technology from Donghua University, China. His research interest lies at the intersection of human-computer interaction and computer vision.
\end{IEEEbiography}

\begin{IEEEbiography}[{\includegraphics[width=1in,height=1.25in,clip,keepaspectratio]{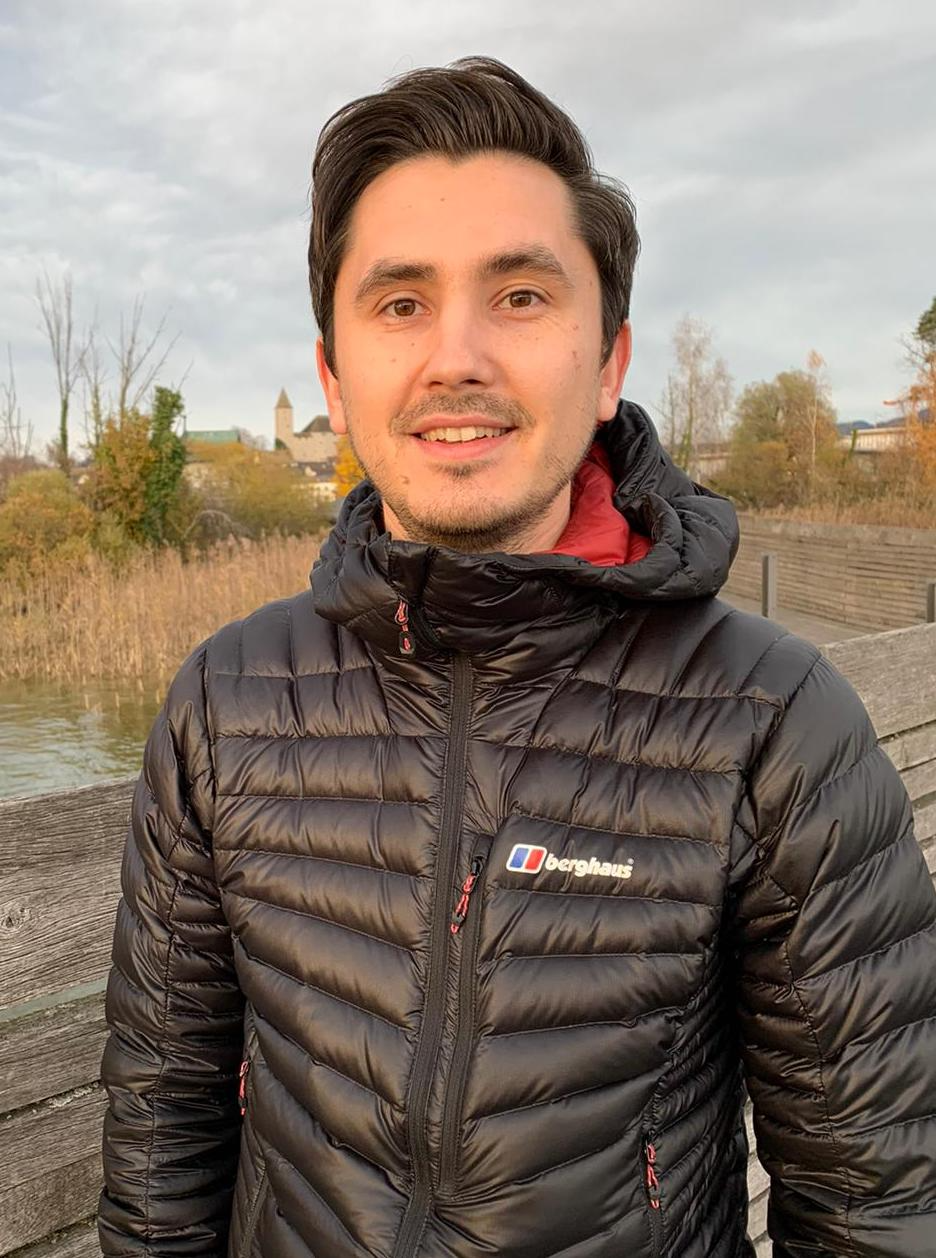}}]{Mihai~B\^{a}ce} is a post-doctoral researcher in the Perceptual User Interfaces group at the University of Stuttgart, Germany. He did his PhD at ETH Zurich, Switzerland, at the Institute for Intelligent Interactive Systems. He received his MSc. in Computer Science from École Polytechnique Fédérale de Lausanne, Switzerland, and his BSc. in Computer Science from the Technical University of Cluj-Napoca, Romania. His research interests include computational human-computer interaction with a focus on sensing and modelling user attention.
\end{IEEEbiography}

\begin{IEEEbiography}[{\includegraphics[width=1in,height=1.25in,clip,keepaspectratio]{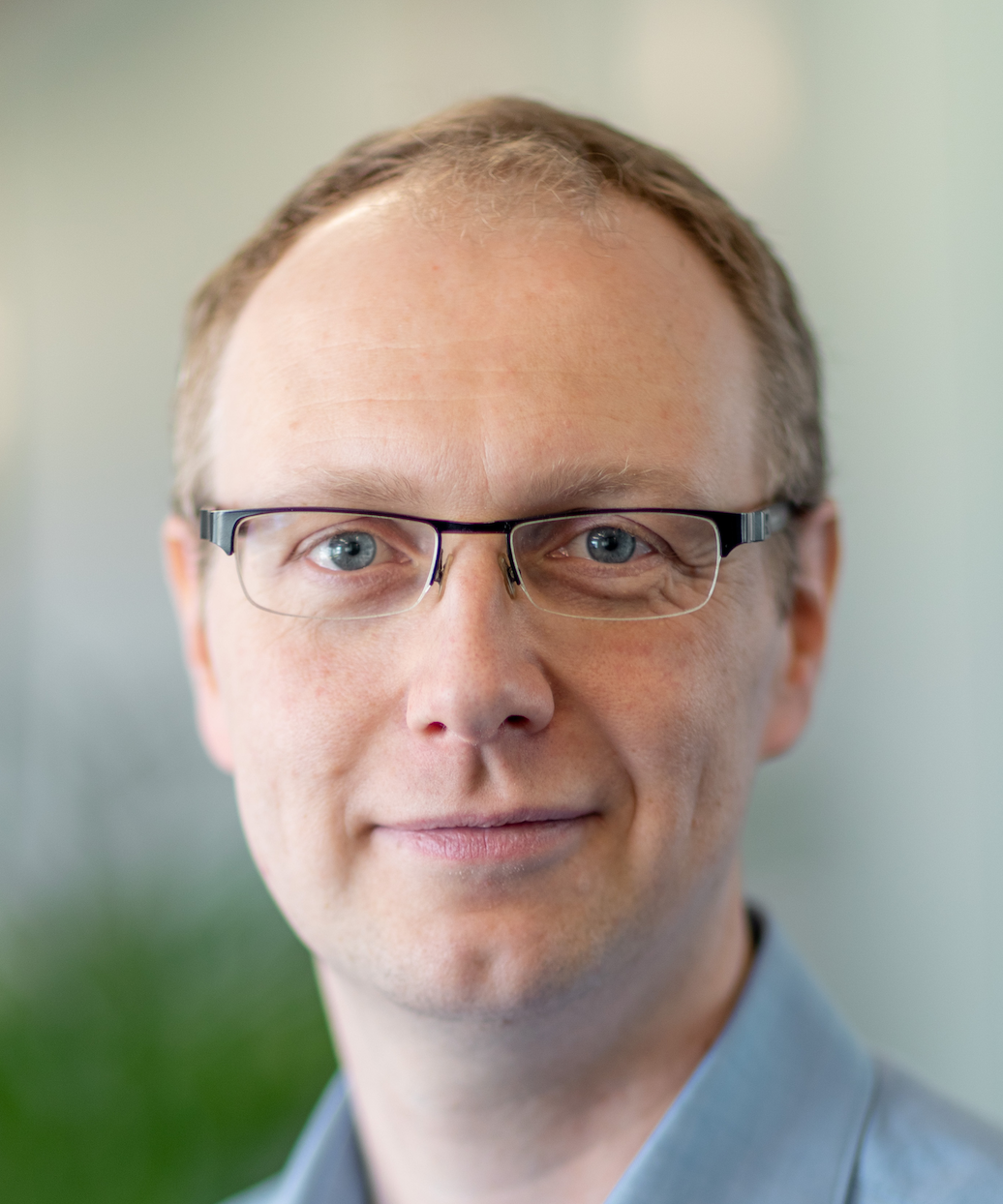}}]{Andreas~Bulling}
is Full Professor of Computer Science at the University of Stuttgart, Germany, where he directs the research group "Human-Computer Interaction and Cognitive Systems". He received his MSc. in Computer Science from the Karlsruhe Institute of Technology, Germany, in 2006 and his PhD in Information Technology and Electrical Engineering from ETH Zurich, Switzerland, in 2010. Before, Andreas Bulling was a Feodor Lynen and Marie Curie Research Fellow at the University of Cambridge, UK, and a Senior Researcher at the Max Planck Institute for Informatics, Germany. His research interests include computer vision, machine learning, and human-computer interaction.
\end{IEEEbiography}

\vfill

\enlargethispage{-5in}

\end{document}

\title{Supplementary Materials for \\``VisRecall: Quantifying information Visualisation Recallability via Question Answering''}

%\author{Yao~Wang,~Sruthi~Radhakrishnan,~Mihai~B\^{a}ce,~and~Andreas~Bulling 
\author{Yao~Wang,~Chuhan~Jiao,~Mihai~B\^{a}ce,~and~Andreas~Bulling% <-this % stops a space
%\IEEEcompsocitemizethanks{\IEEEcompsocthanksitem 
%Yao~Wang,~Sruthi~Radhakrishnan,~Mihai~B\^{a}ce,~and~Andreas~Bulling are with the
%Yao~Wang,~Mihai~B\^{a}ce,~and~Andreas~Bulling are with the Institute for Visualisation and Interactive Systems, University of Stuttgart, Germany \protect\\
% note need leading \protect in front of \\ to get a newline within \thanks as
% \\ is fragile and will error, could use \hfil\break instead.
%E-mail: \{yao.wang, mihai.bace, andreas.bulling\}@vis.uni-stuttgart.de
%, st170459@stud.uni-stuttgart.de
%\IEEEcompsocthanksitem Yao Wang is the corresponding author.}% <-this % stops an unwanted space
%\thanks{Manuscript received xx xx, 2021; revised xx xx, 202x.}
}

% note the % following the last \IEEEmembership and also \thanks - 
% these prevent an unwanted space from occurring between the last author name
% and the end of the author line. i.e., if you had this:
% 
% \author{....lastname \thanks{...} \thanks{...} }
%                     ^------------^------------^----Do not want these spaces!
%
% a space would be appended to the last name and could cause every name on that
% line to be shifted left slightly. This is one of those "LaTeX things". For
% instance, "\textbf{A} \textbf{B}" will typeset as "A B" not "AB". To get
% "AB" then you have to do: "\textbf{A}\textbf{B}"
% \thanks is no different in this regard, so shield the last } of each \thanks
% that ends a line with a % and do not let a space in before the next \thanks.
% Spaces after \IEEEmembership other than the last one are OK (and needed) as
% you are supposed to have spaces between the names. For what it is worth,
% this is a minor point as most people would not even notice if the said evil
% space somehow managed to creep in.

% The paper headers
%\markboth{IEEE Transactions on Vizualisation and Computer Graphics}%
\markboth{}%
{Wang \MakeLowercase{\textit{et al.}}: VisRecall: Quantifying Information Visualisation Recallability via Question Answering}
\maketitle

% To allow for easy dual compilation without having to reenter the
% abstract/keywords data, the \IEEEtitleabstractindextext text will
% not be used in maketitle, but will appear (i.e., to be "transported")
% here as \IEEEdisplaynontitleabstractindextext when the compsoc 
% or transmag modes are not selected <OR> if conference mode is selected 
% - because all conference papers position the abstract like regular
% papers do.
\IEEEdisplaynontitleabstractindextext
% \IEEEdisplaynontitleabstractindextext has no effect when using
% compsoc or transmag under a non-conference mode.

% For peer review papers, you can put extra information on the cover
% page as needed:
% \ifCLASSOPTIONpeerreview
% \begin{center} \bfseries EDICS Category: 3-BBND \end{center}
% \fi
%
% For peerreview papers, this IEEEtran command inserts a page break and
% creates the second title. It will be ignored for other modes.
\IEEEpeerreviewmaketitle

%\IEEEraisesectionheading{\section{Introduction}\label{sec:introduction}}
% Computer Society journal (but not conference!) papers do something unusual
% with the very first section heading (almost always called "Introduction").
% They place it ABOVE the main text! IEEEtran.cls does not automatically do
% this for you, but you can achieve this effect with the provided
% \IEEEraisesectionheading{} command. Note the need to keep any \label that
% is to refer to the section immediately after \section in the above as
% \IEEEraisesectionheading puts \section within a raised box.

% if have a single appendix:
%\appendix[Proof of the Zonklar Equations]
% or
%\appendix  % for no appendix heading
% do not use \section anymore after \appendix, only \section*
% is possibly needed

% use appendices with more than one appendix
% then use \section to start each appendix
% you must declare a \section before using any
% \subsection or using \label (\appendices by itself
% starts a section numbered zero.)
%

% Can use something like this to put references on a page
% by themselves when using endfloat and the captionsoff option.
\ifCLASSOPTIONcaptionsoff
  \newpage
\fi

% trigger a \newpage just before the given reference
% number - used to balance the columns on the last page
% adjust value as needed - may need to be readjusted if
% the document is modified later
%\IEEEtriggeratref{8}
% The "triggered" command can be changed if desired:
%\IEEEtriggercmd{\enlargethispage{-5in}}

% references section

% can use a bibliography generated by BibTeX as a .bbl file
% BibTeX documentation can be easily obtained at:
% http://mirror.ctan.org/biblio/bibtex/contrib/doc/
% The IEEEtran BibTeX style support page is at:
% http://www.michaelshell.org/tex/ieeetran/bibtex/
%\bibliographystyle{IEEEtran}
% argument is your BibTeX string definitions and bibliography database(s)
%\bibliography{IEEEabrv,../bib/paper}
%
% <OR> manually copy in the resultant .bbl file
% set second argument of \begin to the number of references
% (used to reserve space for the reference number labels box)

\begin{figure}[h]
    \centering
    \includegraphics[width=1.2\columnwidth]{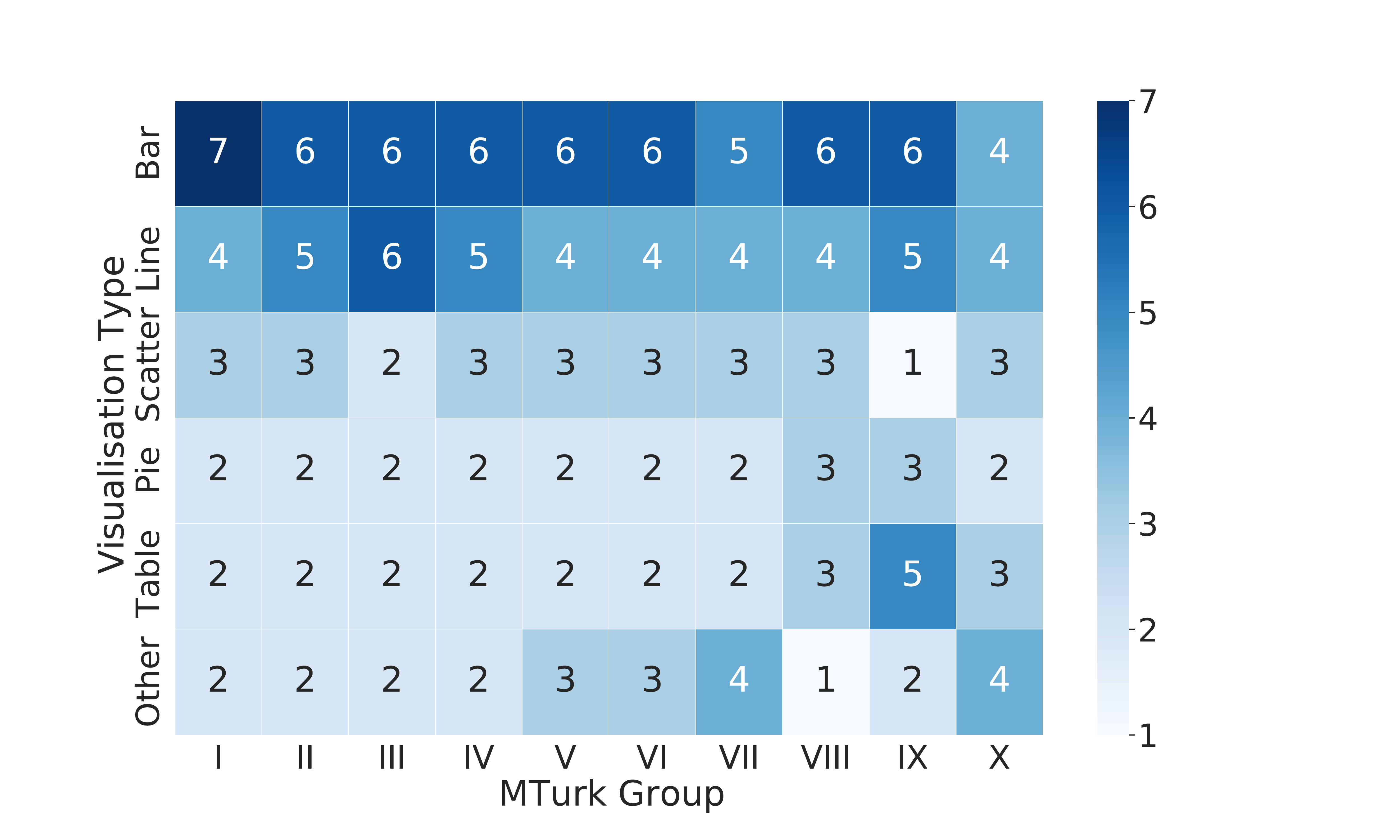}
    \caption{Visualisation type distribution, i.e. the number of visualisations of a certain type, among ten MTurk groups.}
    \label{fig:vistype_amt}
\end{figure}

\begin{figure*}[ht]
    \centering
    \includegraphics[width=\textwidth]{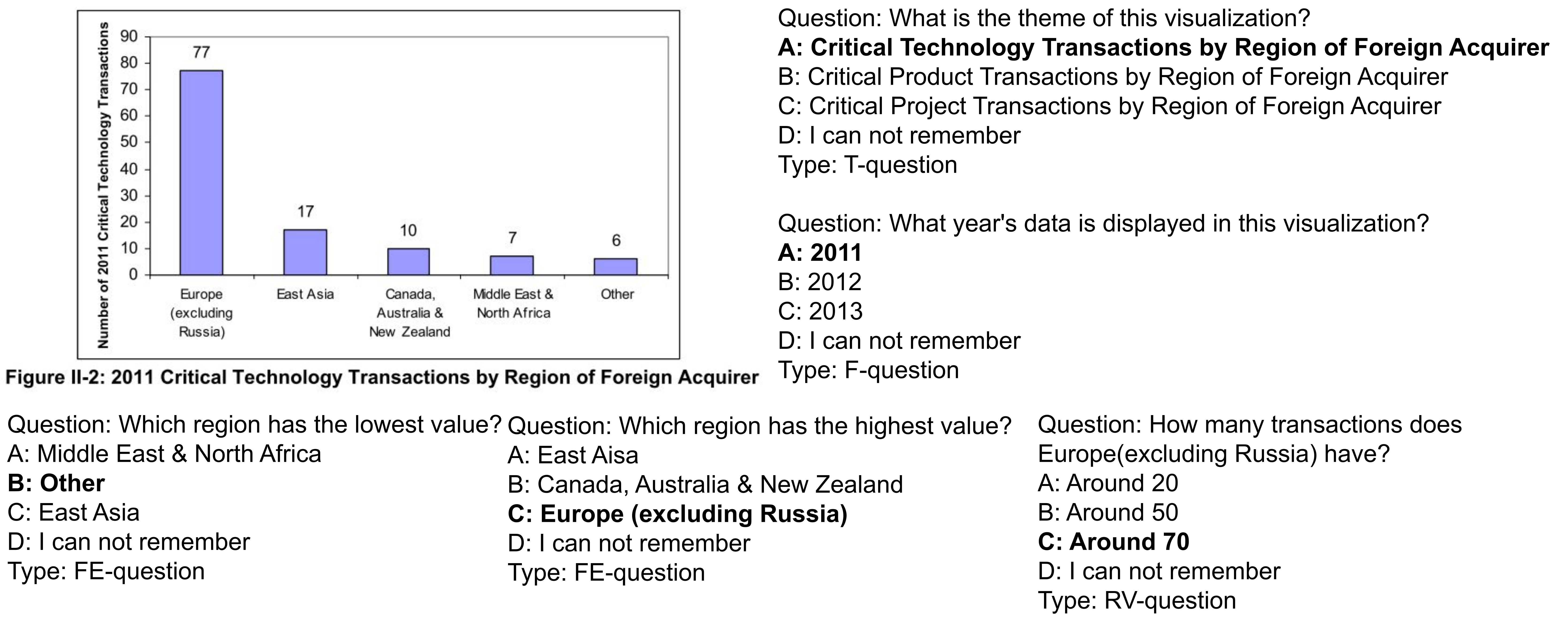}
    \caption{Example visualisation of bar plots from \datasetNameShort dataset with five multiple-choice questions. The correct answer to each question is shown in \textbf{bold}.}
    \label{fig:visrec_bar}
\end{figure*}

\begin{figure*}[ht]
    \centering
    \includegraphics[width=\textwidth]{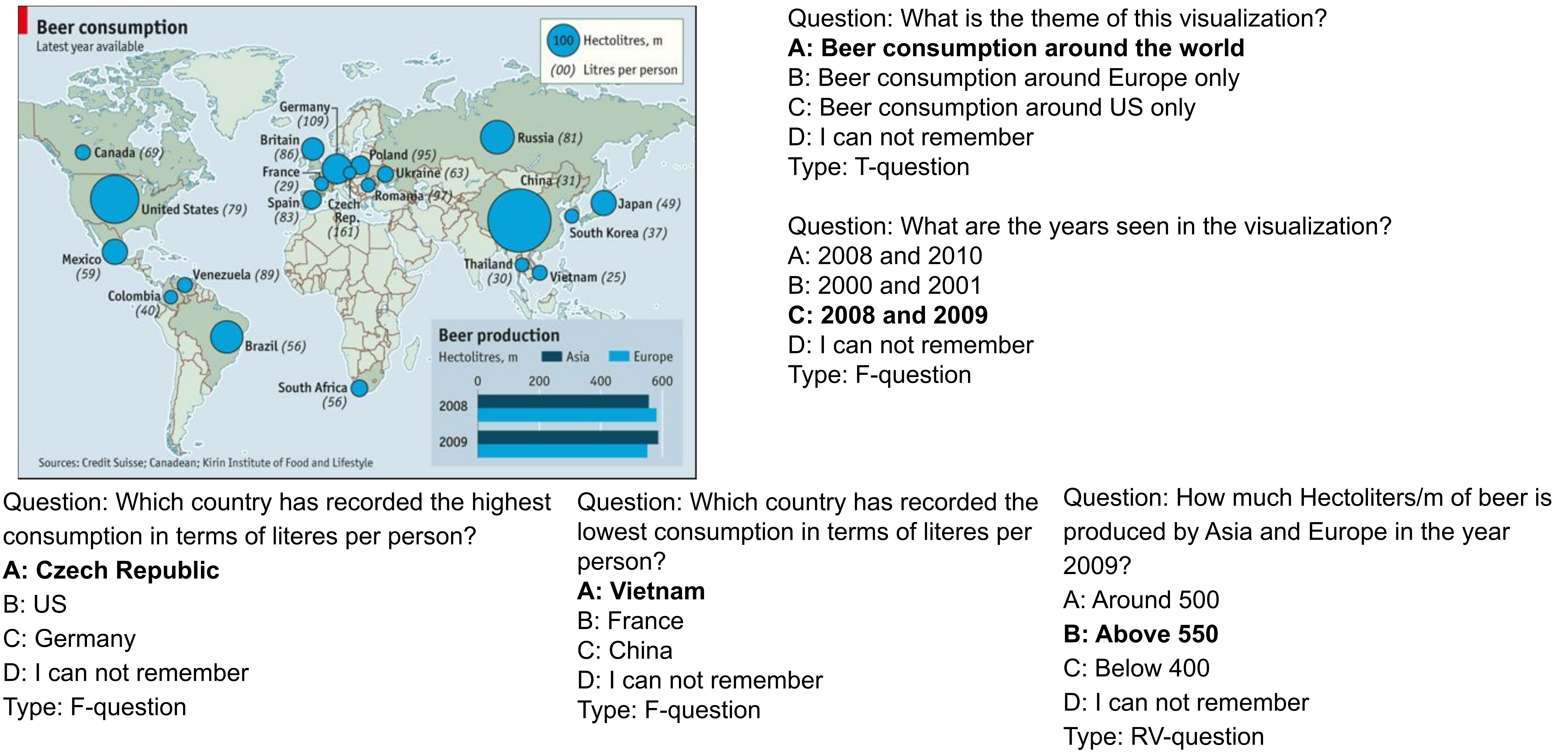}
    \caption{Example visualisation of \textit{others} from \datasetNameShort dataset with five multiple-choice questions. The correct answer to each question is shown in \textbf{bold}.}
    \label{fig:visrec_other}
\end{figure*}

%\appendices
%\section{}
This document contains the visualisation type distribution among MTurk groups (Figure~\ref{fig:vistype_amt}), and additional examples from the \datasetNameShort dataset for each visualisation type~(Figure~\ref{fig:visrec_bar}\,-\,\ref{fig:visrec_other}). Moreover, we provide the \cmr{hit rate, false alarm rate,} memorability~(recognisability) and recallability scores of all visualisations in each quadrant in Figure \cmr{9}~(Right) from the main manuscript~(Tables~\ref{table:topleft}, \ref{table:topright}, and \ref{table:bottomleft}).

\begin{figure*}[ht]
    \centering
    \includegraphics[width=\textwidth]{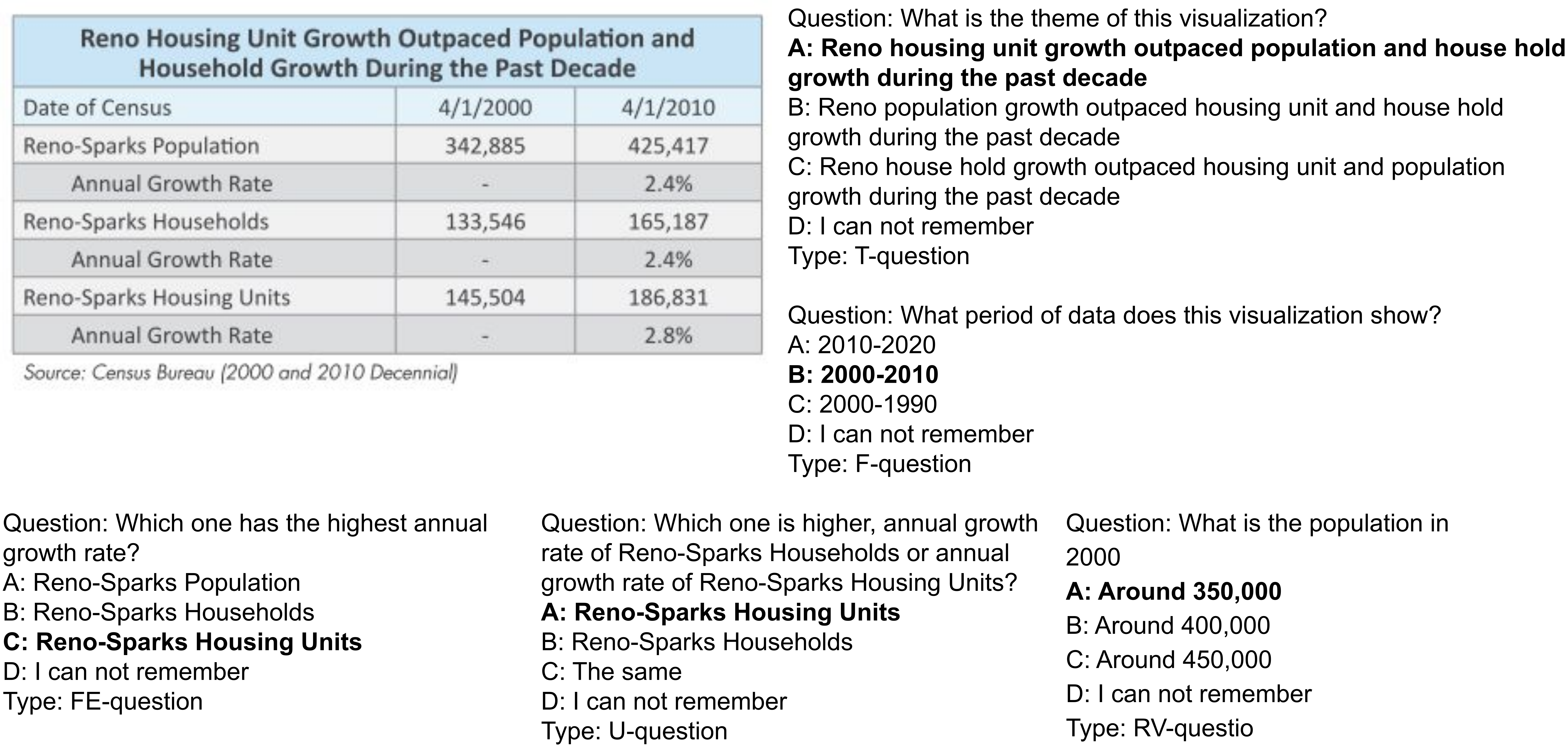}
    \caption{Example visualisation of tables from \datasetNameShort dataset with five multiple-choice questions. The correct answer to each question is shown in \textbf{bold}.}
    \label{fig:visrec_table}
\end{figure*}

\begin{figure*}[ht]
    \centering
    \includegraphics[width=\textwidth]{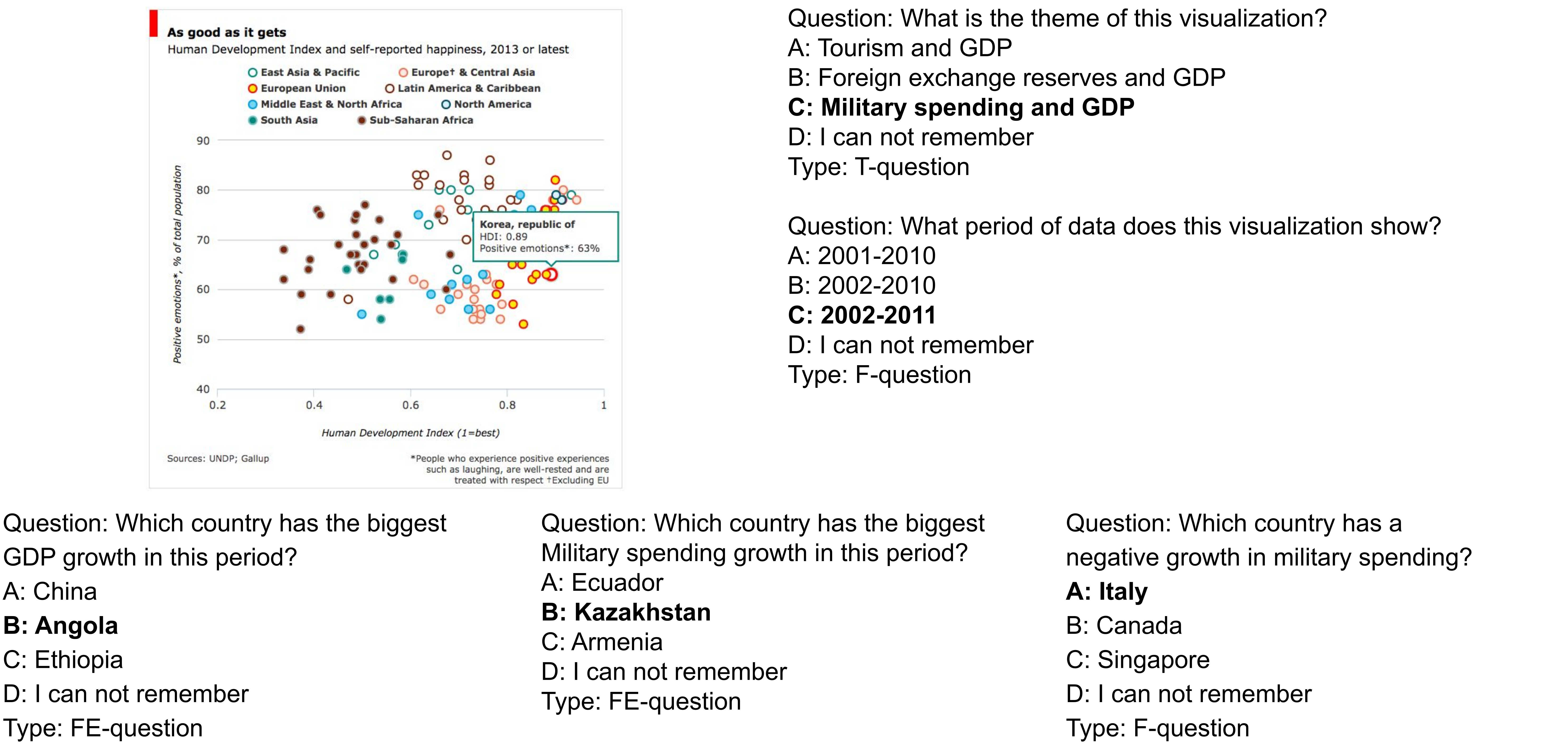}
    \caption{Example visualisation of scatter plots from \datasetNameShort dataset with five multiple-choice questions. The correct answer to each question is shown in \textbf{bold}.}
    \label{fig:visrec_scatter}
\end{figure*}

\begin{figure*}[ht]
    \centering
    \includegraphics[width=\textwidth]{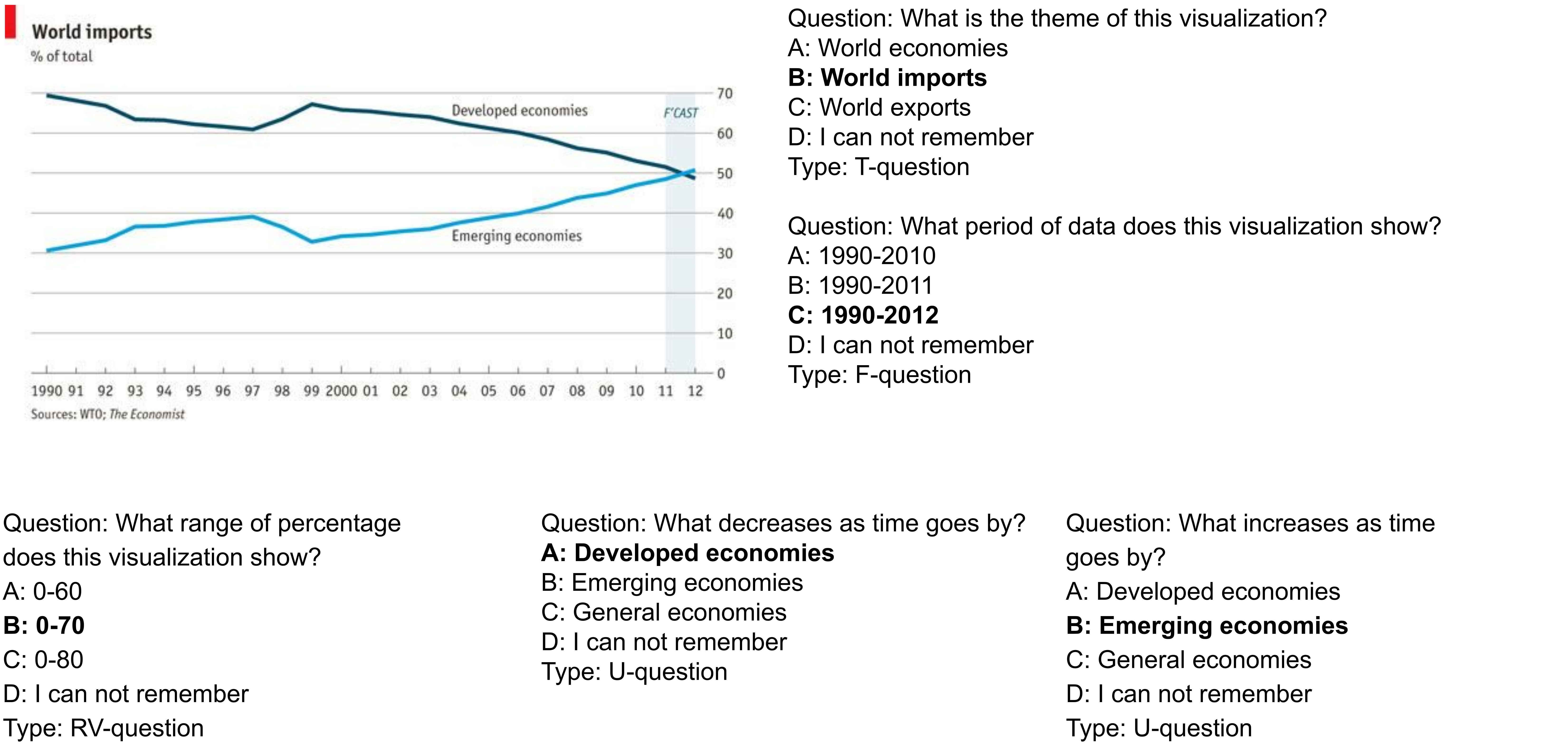}
    \caption{Example visualisation of line plots from \datasetNameShort dataset with five multiple-choice questions. The correct answer to each question is shown in \textbf{bold}.}
    \label{fig:visrec_line}
\end{figure*}

\begin{figure*}[ht]
    \centering
    \includegraphics[width=\textwidth]{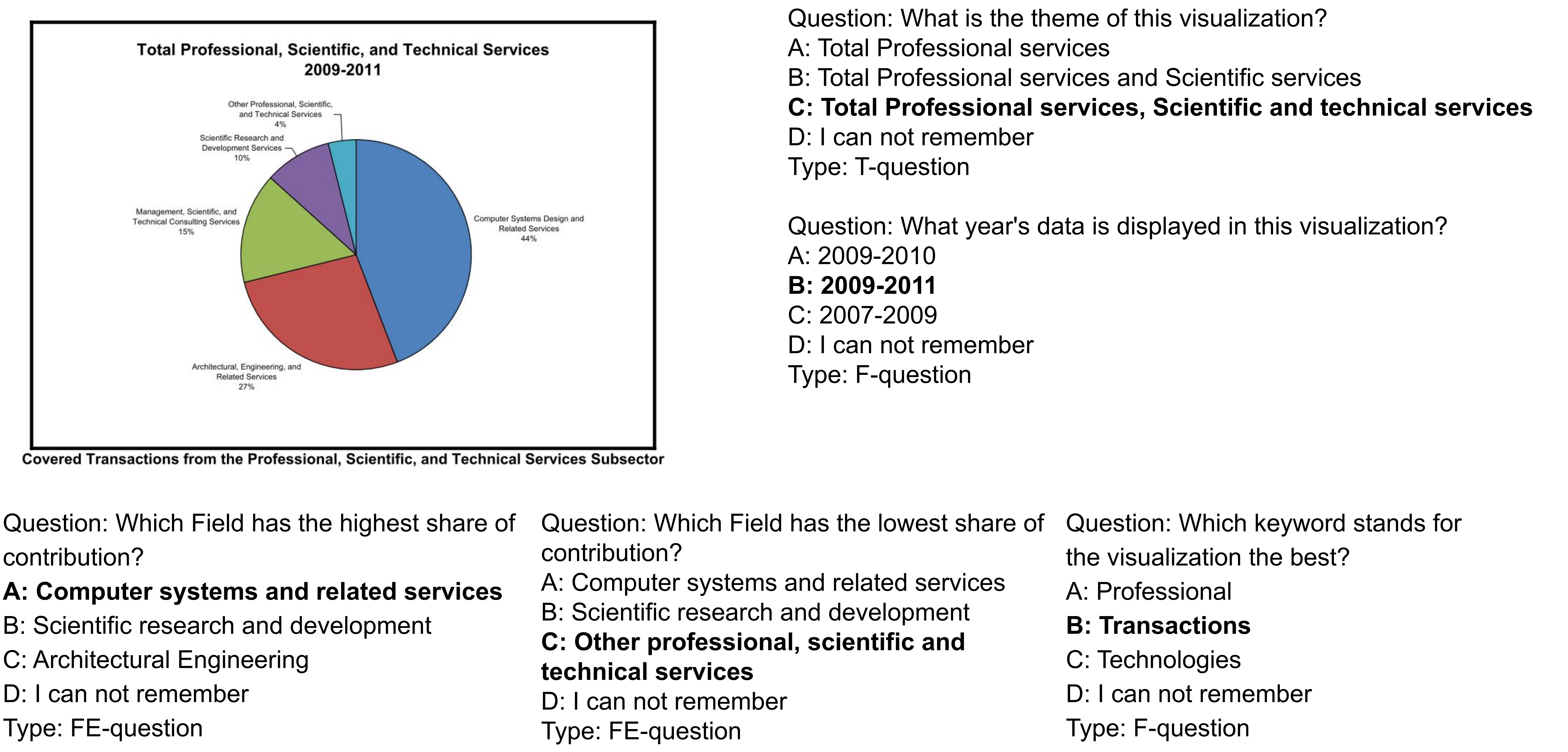}
    \caption{Example visualisation of pie plots from \datasetNameShort dataset with five multiple-choice questions. The correct answer to each question is shown in \textbf{bold}.}
    \label{fig:visrec_pie}
\end{figure*}

\begin{table*}[t]
  \centering
  \begin{tabular}{ | c | l | l | l | l | }
    \hline
    \textbf{Visualisations} & \textbf{Hit Rate} & \textbf{False Alarm Rate} & \textbf{Memorability (Recognisability)} & \textbf{Recallability} \\ \hline
    \begin{minipage}[b]{0.7\columnwidth}
		\centering
		\raisebox{-.5\height}{\includegraphics[width=\linewidth]{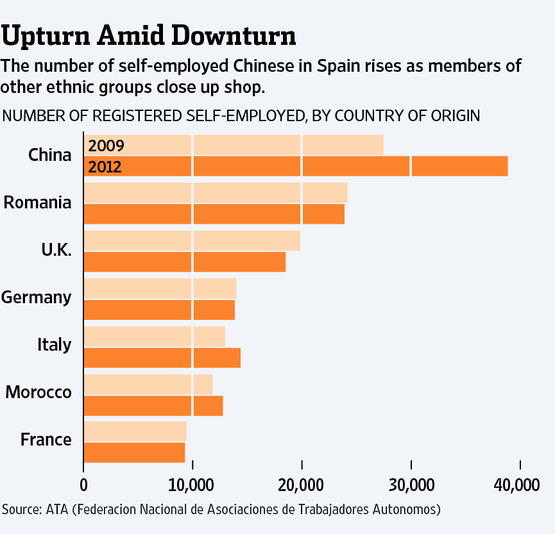}}
	\end{minipage}
    & 1 & 0.026 & 3.986 & 0.358
    \\ \hline
    \begin{minipage}[b]{0.7\columnwidth}
		\centering
		\raisebox{-.5\height}{\includegraphics[width=\linewidth]{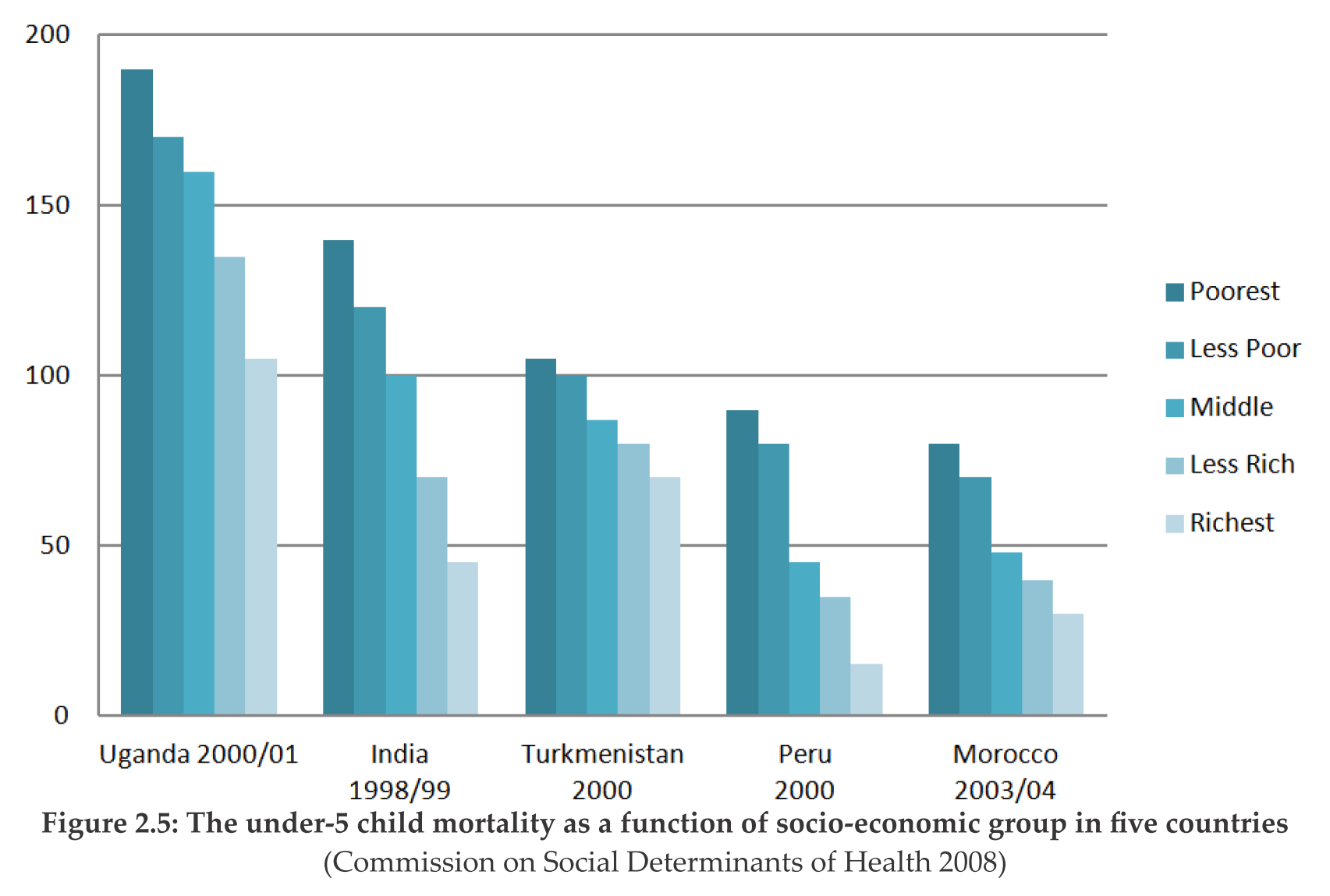}}
	\end{minipage}
    & 1 & 0.051 & 3.669 & 0.183
    \\ \hline
    \begin{minipage}[b]{0.7\columnwidth}
		\centering
		\raisebox{-.5\height}{\includegraphics[width=\linewidth]{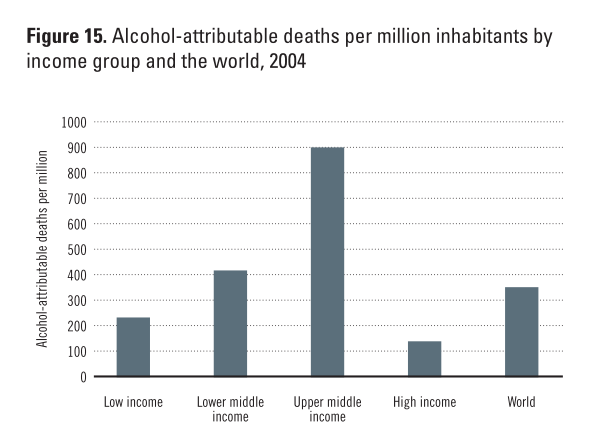}}
	\end{minipage}
    & 1 & 0.077 & 3.463 & 0.075
    \\ \hline
    \begin{minipage}[b]{0.7\columnwidth}
		\centering
		\raisebox{-.5\height}{\includegraphics[width=\linewidth]{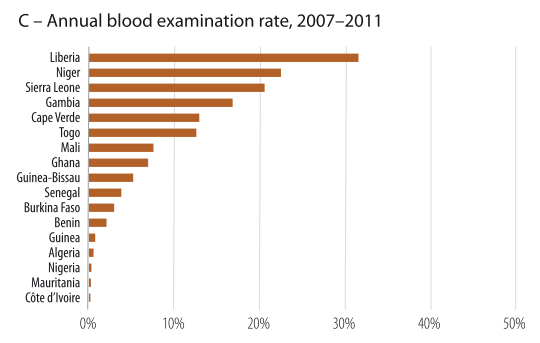}}
	\end{minipage}
    & 0.875 & 0 & 3.382 & 0.292
    \\ \hline
  \end{tabular}
  \caption{\cmr{Hit rate, false alarm rate,} memorability (recognisability) and recallability scores of all visualisations in top-left quadrant in Figure \cmr{9} (Right) from the main manuscript.}
    \label{table:topleft}
\end{table*}

\begin{table*}[t]
  \centering
  \begin{tabular}{ | c | l | l | l | l | }
    \hline
    \textbf{Visualisations} & \textbf{Hit Rate} & \textbf{False Alarm Rate} & \textbf{Memorability (Recognisability)} & \textbf{Recallability} \\ \hline
    \begin{minipage}[b]{0.66\columnwidth}
		\centering
		\raisebox{-.5\height}{\includegraphics[width=\linewidth]{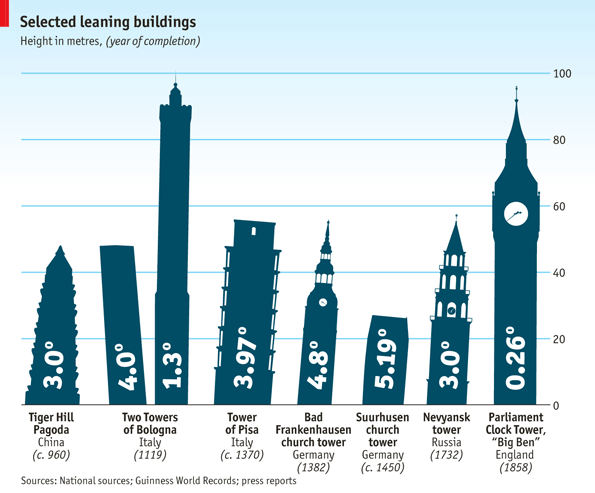}}
	\end{minipage}
    & 1 & 0 & 4.268 & 0.633
    \\ \hline
    \begin{minipage}[b]{0.66\columnwidth}
		\centering
		\raisebox{-.5\height}{\includegraphics[width=\linewidth]{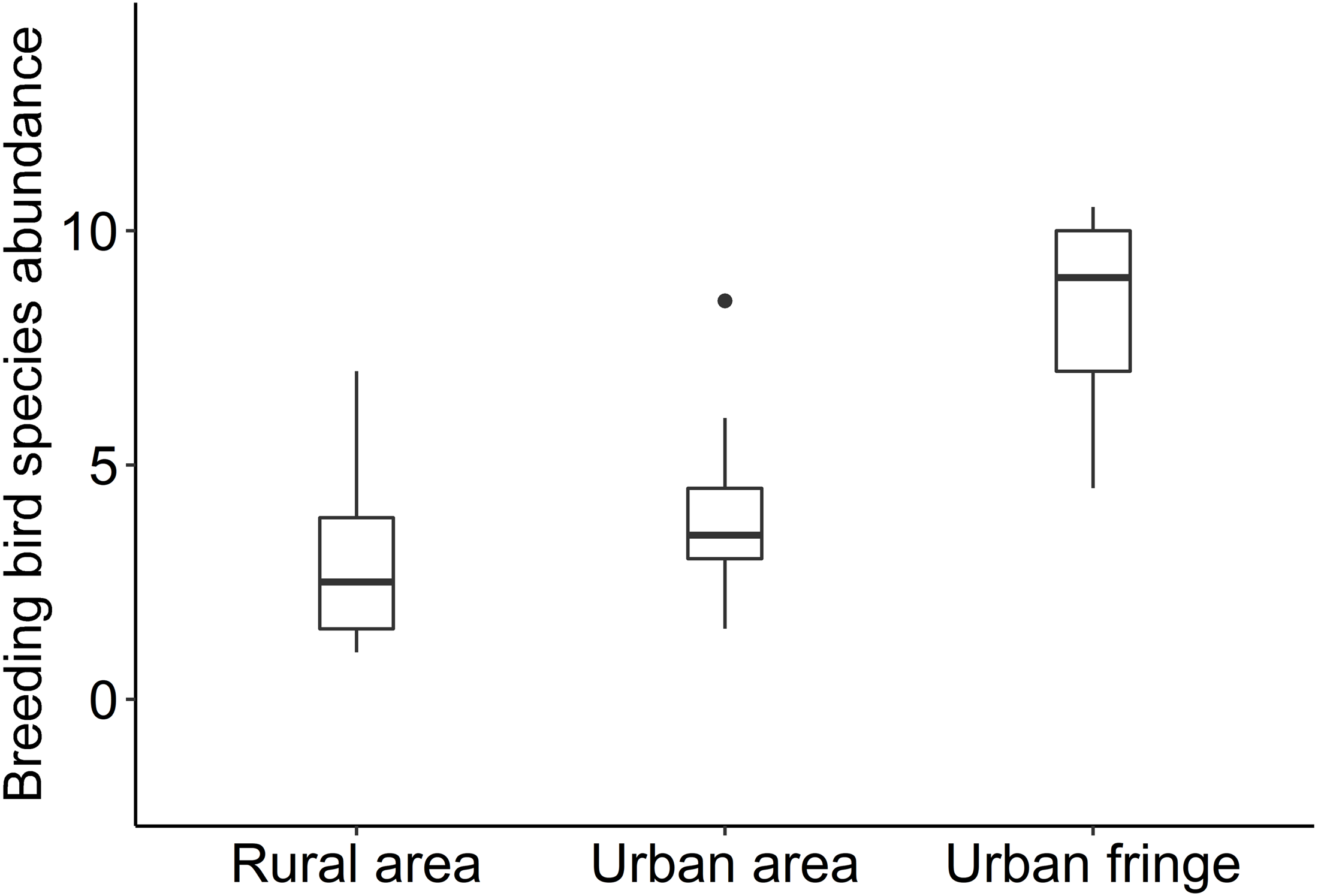}}
	\end{minipage}
    & 0.970 & 0 & 4.203 & 0.679
    \\ \hline
    \begin{minipage}[b]{0.66\columnwidth}
		\centering
		\raisebox{-.5\height}{\includegraphics[width=\linewidth]{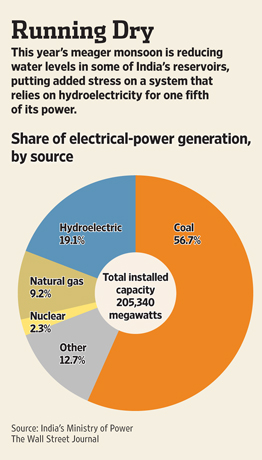}}
	\end{minipage}
    & 1 & 0.026 & 3.986 & 0.658
    \\ \hline
    \begin{minipage}[b]{0.66\columnwidth}
		\centering
		\raisebox{-.5\height}{\includegraphics[width=\linewidth]{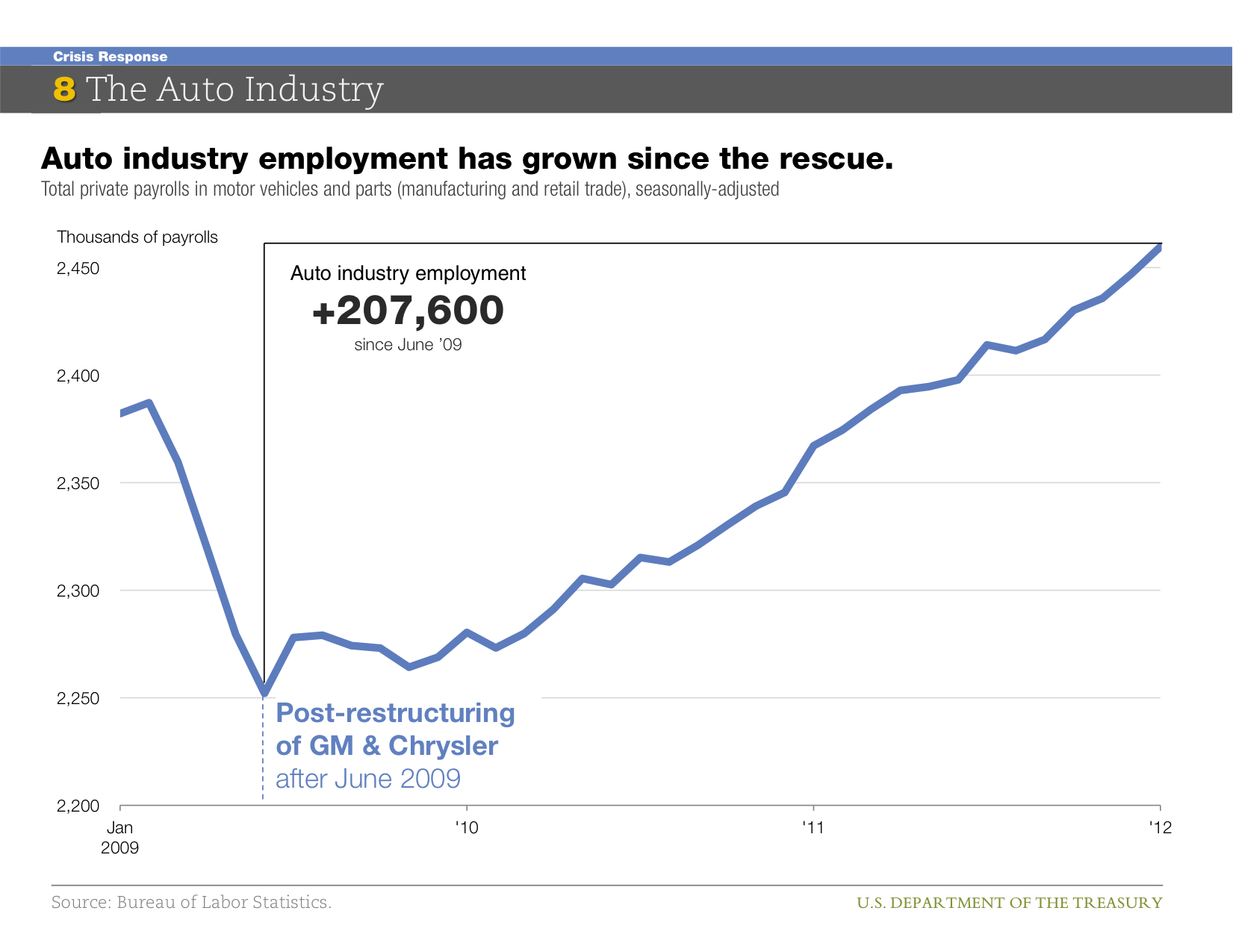}}
	\end{minipage}
    & 1 & 0.026 & 3.986 & 0.650
    \\ \hline
  \end{tabular}
  \caption{\cmr{Hit rate, false alarm rate,} memorability (recognisability) and recallability scores of all visualisations in top-right quadrant in Figure \cmr{9} (Right) from the main manuscript.}
    \label{table:topright}
\end{table*}

\begin{table*}[t]
  \centering
  \begin{tabular}{ | c | l | l | l | l | }
    \hline
    \textbf{Visualisations} & \textbf{Hit Rate} & \textbf{False Alarm Rate} & \textbf{Memorability (Recognisability)} & \textbf{Recallability} \\ \hline
    \begin{minipage}[b]{0.7\columnwidth}
		\centering
		\raisebox{-.5\height}{\includegraphics[width=\linewidth]{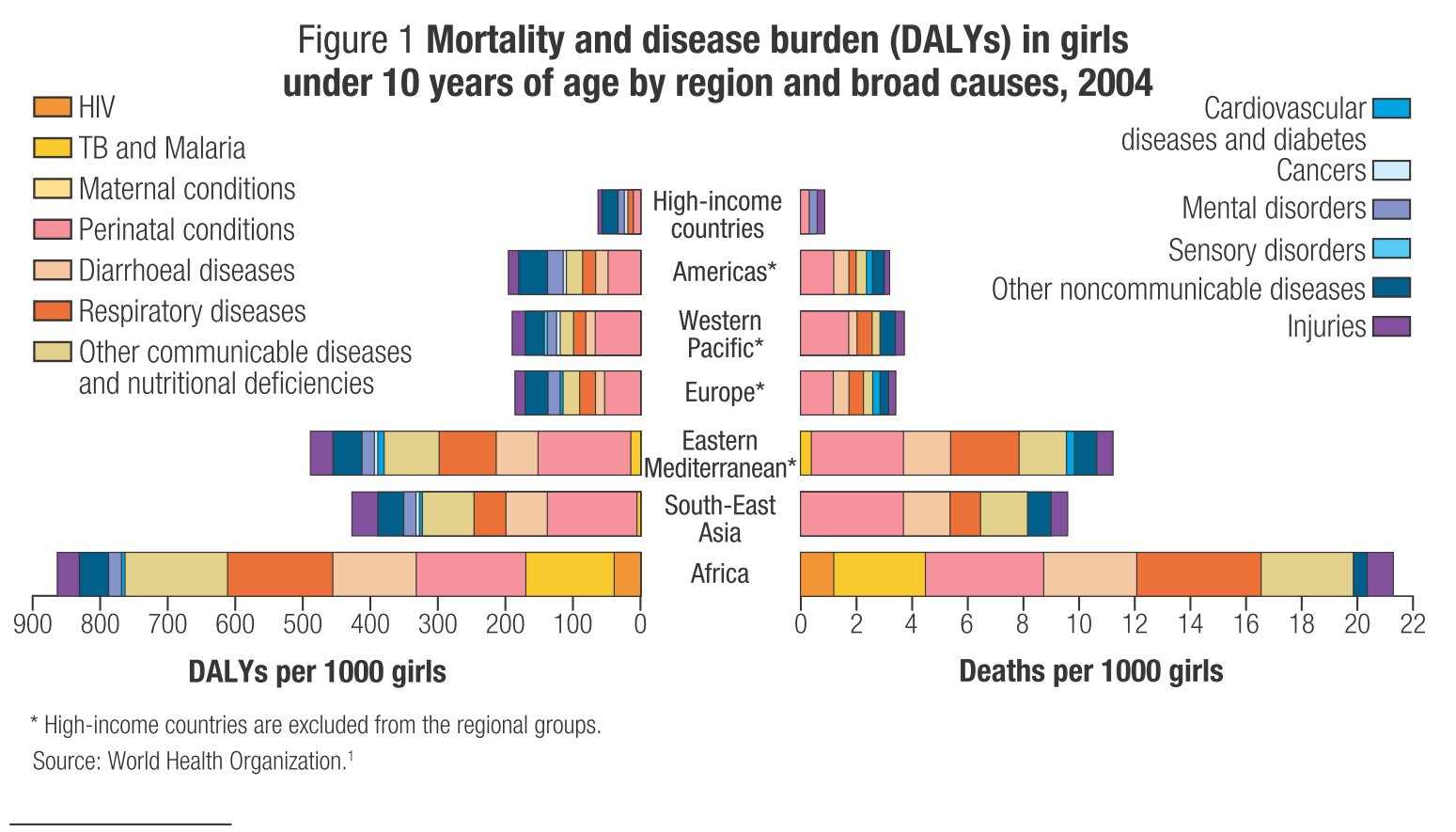}}
	\end{minipage}
    & 0.923 & 0.833 & 0.459 & 0.318
    \\ \hline
    \begin{minipage}[b]{0.7\columnwidth}
		\centering
		\raisebox{-.5\height}{\includegraphics[width=\linewidth]{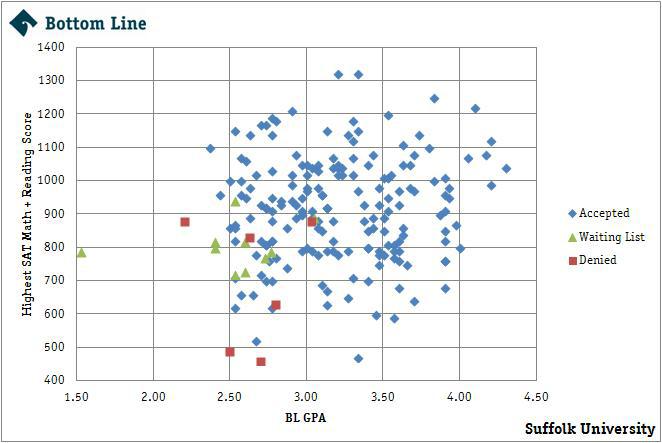}}
	\end{minipage}
    & 0.75 & 0.429 & 0.855 & 0.317
    \\ \hline
    \begin{minipage}[b]{0.7\columnwidth}
		\centering
		\raisebox{-.5\height}{\includegraphics[width=\linewidth]{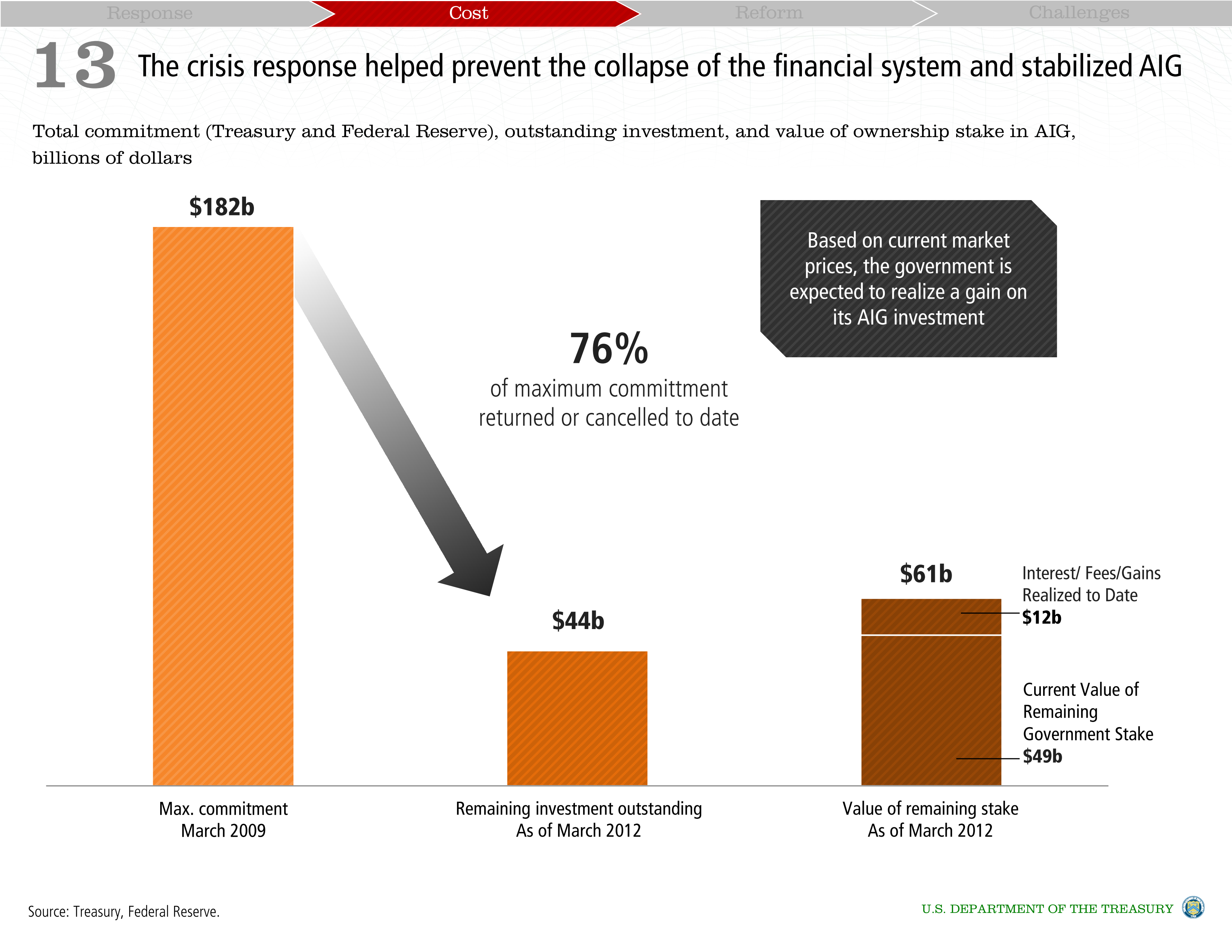}}
	\end{minipage}
    & 0.717 & 0.286 & 1.139 & 0.337
    \\ \hline
    \begin{minipage}[b]{0.7\columnwidth}
		\centering
		\raisebox{-.5\height}{\includegraphics[width=\linewidth]{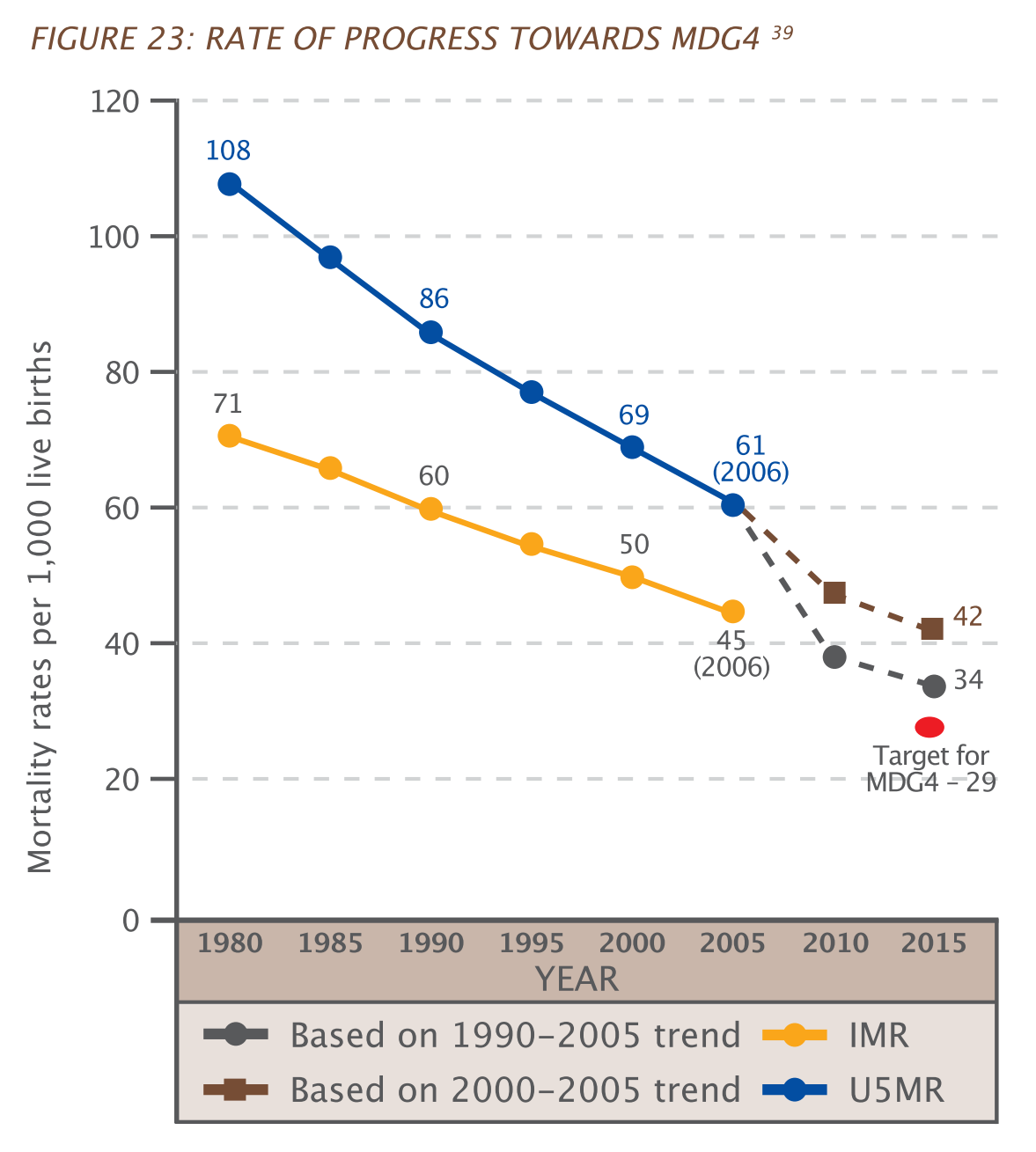}}
	\end{minipage}
    & 0.783 & 0.393 & 1.055 & 0.323
    \\ \hline
  \end{tabular}
  \caption{\cmr{Hit rate, false alarm rate,} memorability (recognisability) and recallability scores of all visualisations in bottom-left quadrant in Figure \cmr{9} (Right) from the main manuscript.}
    \label{table:bottomleft}
\end{table*}

%\clearpage
%\bibliographystyle{IEEEtranN}
%\bibliography{references}

% biography section
% 
% If you have an EPS/PDF photo (graphicx package needed) extra braces are
% needed around the contents of the optional argument to biography to prevent
% the LaTeX parser from getting confused when it sees the complicated
% \includegraphics command within an optional argument. (You could create
% your own custom macro containing the \includegraphics command to make things
% simpler here.)
%\begin{IEEEbiography}[{\includegraphics[width=1in,height=1.25in,clip,keepaspectratio]{mshell}}]{Michael Shell}
% or if you just want to reserve a space for a photo:

% You can push biographies down or up by placing
% a \vfill before or after them. The appropriate
% use of \vfill depends on what kind of text is
% on the last page and whether or not the columns
% are being equalized.

%\vfill

% Can be used to pull up biographies so that the bottom of the last one
% is flush with the other column.
%\enlargethispage{-5in}

% that's all folks